\documentclass{emulateapj}

\slugcomment{Accepted version from May 29, 2013}
\shorttitle{The Magellanic Stream toward Fairall\,9}
\shortauthors{P. Richter et al.}

\begin{document}

\title{
The COS/UVES Survey of the Magellanic Stream: \\
II. Evidence for a complex enrichment history of the Stream 
from the Fairall\,9 sightline}

\author{Philipp Richter\altaffilmark{1,2},
Andrew J. Fox\altaffilmark{3},
Bart P. Wakker\altaffilmark{4},
Nicolas Lehner\altaffilmark{5},
J. Christopher Howk\altaffilmark{5},
Joss Bland-Hawthorn\altaffilmark{6},
Nadya Ben Bekhti\altaffilmark{7},
\and
Cora Fechner\altaffilmark{1}}

\affil{$^1$Institut f\"ur Physik und Astronomie, Universit\"at Potsdam,
Haus 28, Karl-Liebknecht-Str.\,24/25, 14476 Golm (Potsdam),
Germany}
\affil{$^2$Leibniz-Institut f\"ur Astrophysik Potsdam (AIP),
An der Sternwarte 16, 14482 Potsdam, Germany}
\affil{$^3$Space Telescope Science Institute, Baltimore, MD 21218, USA}
\affil{$^4$Department of Astronomy, University of Wisconsin-Madison,
475 N. Charter Street, Madison, WI\,53706, USA}
\affil{$^5$Department of Physics, University of Notre Dame, 225 Nieuwland Science Hall,
Notre Dame, IN 46556, USA}
\affil{$^6$Institute of Astronomy, School of Physics, University of Sydney,
NSW 2006, Australia}
\affil{$^7$Argelander-Institut f\"ur Astronomie, Universit\"at Bonn,
Auf dem H\"ugel 71, 53121 Bonn, Germany}

%

\begin{abstract}
We present a multi-wavelength
\footnote{Based on observations obtained with the NASA/ESA
Hubble Space Telescope, which is operated by the Space Telescope Science
Institute (STScI) for the Association of Universities for Research in Astronomy,
Inc., under NASA contract NAS5D26555, and
on observations collected at the European Organisation for
Astronomical Research in the Southern Hemisphere, Chile 
under Program ID\,085.C$-$0172(A).}
study of the Magellanic Stream (MS), a massive gaseous
structure in the Local Group that is believed to represent material stripped
from the Magellanic Clouds. We use ultraviolet, optical and radio data
obtained with \emph{HST}/COS, VLT/UVES, \emph{FUSE}, GASS, and ATCA to study
metal abundances and physical conditions in the Stream toward the
quasar Fairall\,9. Line absorption in the MS from a large number of metal 
ions and from molecular hydrogen is detected in up to seven absorption components,
indicating the presence of multi-phase gas.
From the analysis of unsaturated S\,{\sc ii} absorption, in combination
with a detailed photoionization model, we obtain a surprisingly
high $\alpha$ abundance in the Stream toward Fairall\,9 of
[S/H$]=-0.30\pm0.04$ ($0.50$ solar). This value is 5 times
higher than what is found along other MS sightlines based on
similar COS/UVES data sets. In contrast, the measured nitrogen
abundance is found to be substantially lower ([N/H$]=-1.15\pm0.06$),
implying a very low [N/$\alpha$] ratio of $-0.85$ dex. 
The substantial differences in the chemical composition of MS 
toward Fairall\,9 compared to other sightlines point toward a
complex enrichment history of the Stream. We favour a scenario, 
in which the gas toward Fairall\,9 was locally enriched with $\alpha$ elements
by massive stars and then was separated from the Magellanic Clouds
before the delayed nitrogen enrichment from intermediate-mass stars 
could set in. Our results support (but do not require) the idea that there
is a metal-enriched filament in the Stream toward Fairall\,9 that 
originates in the LMC.  
\end{abstract}

%

 
%

\section{Introduction}

The distribution of neutral and ionized gas in the circumgalactic 
environment of galaxies is known to be an important indicator of
the past and present evolution of galaxies. Both the infall of 
metal-poor gas from intergalactic space and from satellite galaxies
and the outflow of metal-rich gaseous material through galactic winds
represent key phenomena that determine the spatial distribution and 
the physical state of the circumgalactic gas around massive galaxies.

From observations and theoretical studies, it is known that galaxy 
interactions between gas-rich galaxies can transport large
amounts of neutral and ionized gas into the circumgalactic environment of 
galaxies. In the local Universe, the most massive of these extended tidal 
gas features can be observed in the 21\,cm line of neutral hydrogen (H\,{\sc i}).
The most prominent nearby example of a tidal gas stream produced by
the interaction of galaxies is the Magellanic Stream (MS), a massive 
($\sim 10^8-10^9\,M_{\sun}$) stream of neutral and ionized gas in 
the outer halo of the Milky Way at a distance of $\sim 50-60$ kpc
(e.g., Wannier \& Wrixon 1972; Gardiner \& Noguchi 1996; Weiner \& Williams 1996; 
Putman et al.\,2003; Br\"uns et al.\,2005; 
Fox et al.\,2005, 2010; Koerwer 2009; Besla et al.\,2007, 2010, 2012). The Magellanic
Stream spans over 200\degr\ on the sky (e.g., Nidever et al.\,2010) and has
a (mean) metallicity that is lower than that of the Milky Way, but 
comparable with the metallicity found in the SMC and LMC 
($0.1-0.5$ solar; Lu et al.\,1994; 
Gibson et al.\,2000; Sembach et al.\,2001; Fox et al.\,2010, 2013). 
The MS also contains dust grains and diffuse molecular hydrogen
(H$_2$; Sembach et al.\,2001; Richter et al.\,2001a). 
A number of theoretical studies, including tidal models and
ram-pressure stripping models, have been carried out to describe the 
Stream's motion in the extended halo of the Milky Way and pinpoint
its origin in one of the two Magellanic Clouds (Gardiner \& Noguchi 1996;
Mastropietro et al.\,2005; Connors et al.\,2006; Besla et al.\,2010; 
Diaz \& Bekki 2011). 

The origin and fate of the Magellanic Stream is 
closely related to the trajectories of LMC and SMC 
(e.g., Connors et al.\,2004, 2006; Besla et al.\,2007), and 
any realistic model of the MS thus needs to consider 
the dynamical and physical state of the 
Milky Way/Magellanic Clouds system as a whole 
(see also Bland-Hawthorn et al.\,2007; Heitsch \& Putman 2009). 
While early tidal models have assumed that the Magellanic Stream is 
a product from the tidal interaction between LMC and SMC as they
periodically orbit the Milky Way (e.g., Gardiner \& Noguchi 1996), 
more recent proper motion measurements of the Magellanic Clouds (MCs)
(Kallivayalil et al.\,2006a, 2006b, 2013) indicate that the MCs may be 
on their first passage around the Milky Way. Some subsequent tidal models
(Besla et al.\,2010; Diaz \& Bekki 2011) thus favour a first-infall
scenario for the Magellanic Stream. Moreover, while many 
models (e.g., Connors et al.\,2006) place the origin 
of the Stream's gaseous material in the SMC, other, more
recent studies trace back at least part of the Stream's gaseous material
in the LMC (e.g., Nidever et al.\,2008). The latter study also highlights 
the role of energetic blowouts from star-forming regions in the LMC
for the formation of the Stream.
Clearly, further theoretical studies and observations are
required to pinpoint the origin of the MS based on different
(independent) methods.

In the first paper in our series analyzing the chemical and physical
conditions in the Magellanic Stream (Fox et al.\,2013; hereafter 
Paper\,I), we have investigated MS absorption in the UV and optical 
along the lines of sight toward RBS\,144, NGC\,7714\, PHL\,2525, and 
HE\,0056$-$3622.
In this paper we analyze the MS using UV and optical absorption-line
spectra of the Seyfert\,1 galaxy Fairall\,9 ($z_{\rm em}=0.047$). 
Located at $l=295.1$ and 
$b=-57.8$ the Fairall\,9 sightline lies only 14.3\degr\ on the sky 
from the SMC. This sightline is the best-studied in absorption of all 
MS directions (Songaila 1981; York et al.\,1982; 
Lu, Savage \& Sembach 1994; Gibson et al.\,2000; 
Richter et al.\,2001a; Sembach et al.\,2003), largely because
the Fairall\,9 is bright in both the optical and the UV and the
Stream's H\,{\sc i} column in this direction is large 
(log $N$(H\,{\sc i}$)\approx 20$; see Gibson et al.\,2000). 
The high column of neutral gas ensures that a wide range of low-ionization
UV metal lines are detectable in the Stream, and even molecular hydrogen 
was observed in the MS toward Fairall\,9 data from the
\emph{Far Ultraviolet Spectroscopic Explorer} (\emph{FUSE}; Richter et al.\,2001a;
Wakker 2006). 
Using a spectrum of Fairall\,9 obtained with the Goddard High Resolution 
Spectrograph (GHRS) onboard the \emph {Hubble Space Telescope} (HST) together 
with Parkes 21\,cm H\,{\sc i} data Gibson et al.\,(2000) derived a metallicity
of the Stream toward Fairall\,9 
of [S/H$]=-0.55\pm0.06^{+0.17}_{-0.21}$ ($\sim 0.3$ solar),
which represented the most accurate metallicity determination of the 
Stream from UV absorption line data at that time. This metallicity is 
consistent with either an SMC or LMC origin of the gas. A difficulty for 
constraining the origin of the MS in one or the other Magellanic Cloud 
arises from the fact that the gas in the Stream was
stripped from its parent galaxy $\sim 1-2$ Gyr ago 
(e.g., Gardiner \& Noguchi 1996; Connors et al.\,2006; Nidever et al.\,2008),
but has not experienced
any further metal enrichment since then, while the parent galaxy 
underwent further chemical evolution. The MS does not contain 
any massive stars (e.g., Mathewson et al.\,1979), in contrast to the 
Magellanic Bridge (Irwin, Kunkel \& Demers 1985). This aspect 
needs to be taken into account when comparing metal abundances 
in the Stream with present-day LMC and SMC abundances.

To increase the accuracy of the metallicity determination of the 
Stream toward Fairall\,9 and to obtain more detailed information on the 
chemical composition of the gas and dust in the Stream, more
accurate spectral data are desirable.
Because the Magellanic Stream is a massive gas cloud with complex
internal kinematics (e.g., Nidever et al.\,2008), data with 
high spectral resolution and a high signal-to-noise (S/N) 
ratio are required to fully resolve the Stream's velocity-component structure and to
detect weak absorption features from the various metal ions that
have their transitions in the ultraviolet (UV) and in the optical. 
As part of our ongoing project to study the properties of the Magellanic Stream 
in absorption along multiple lines of sight (see also Paper\,I)
we obtained high-resolution optical data of Fairall\,9 from 
the Ultraviolet and Visible Echelle Spectrograph (UVES) 
installed on the Very Large Telescope (VLT)
and medium-resolution UV data from the Cosmic Origins Spectrograph 
(COS) onboard the \emph{HST}, both data sets providing
absoption spectra with excellent S/N ratios.
The combination of these data sets, as desribed in this study, 
therefore provides a particular promising strategy
to study in great detail the chemical and physical conditions in the 
Magellanic Stream in this direction.

This paper is organized as follows: in Sect.\,2 we describe the observations
and the data reduction. The column density measurements and the profile
modeling are explained in Sect.\,3. In Sect.\,4 we derive
chemical and physical properties of the gas in the MS. We discuss
our results in Sect.\,5. Finally, a summary of our study is given
in Sect.\,6.

%

\section{Observations and spectral analysis}

\subsection{VLT/UVES observations}

Fairall\,9 was observed with the VLT/UVES spectrograph (Dekker et al.\,2010) 
in 2010 under ESO program ID\,085.C-0172(A)\,(PI: A. Fox). The observations 
were taken in Service Mode using Dichroic\,1 in the 390+580 setting, 
a 0.6\arcsec\ slit, and no rebinning. The observations were
carried out under good seeing conditions ($<$0.8\arcsec).
The raw data were reduced with the standard UVES pipeline, using
calibration frames taken close in time to the corresponding
science frames. The reduction steps involve subtraction of the bias
level, inter-order background, sky background, night sky emission
lines, and cosmic ray hits. The frames were then flat-fielded,
optimally extracted and merged. The wavelength scale was corrected for
atmospheric dispersion and heliocentric velocity and then placed
into the local standard of rest (LSR) velocity frame. Multiple 
exposures on the same target
were registered onto a common wavelength grid and then added.
The final spectra have a very high spectral resolution of 
$R\approx70\,000$ corresponding to a FWHM of $4.3$ km\,s$^{-1}$.
They cover the wavelength range between $3300$ and $6800$ \AA.
The S/N ratio per resolution element is $40$ at
$3500$ \AA\,(Ti\,{\sc ii}), $65$ at $4000$ \AA\,(Ca\,{\sc ii}), 
and $83$ at $6000$ \AA\,(Na\,{\sc i}). The UVES data thus
provide much higher sensitivity and substantially higher 
spectral resolution than previous optical measurements of 
Fairall\,9 (Songaila 1981).
\vspace{0.6cm}

%

\begin{figure}[t!]
\epsscale{1.20}
\plotone{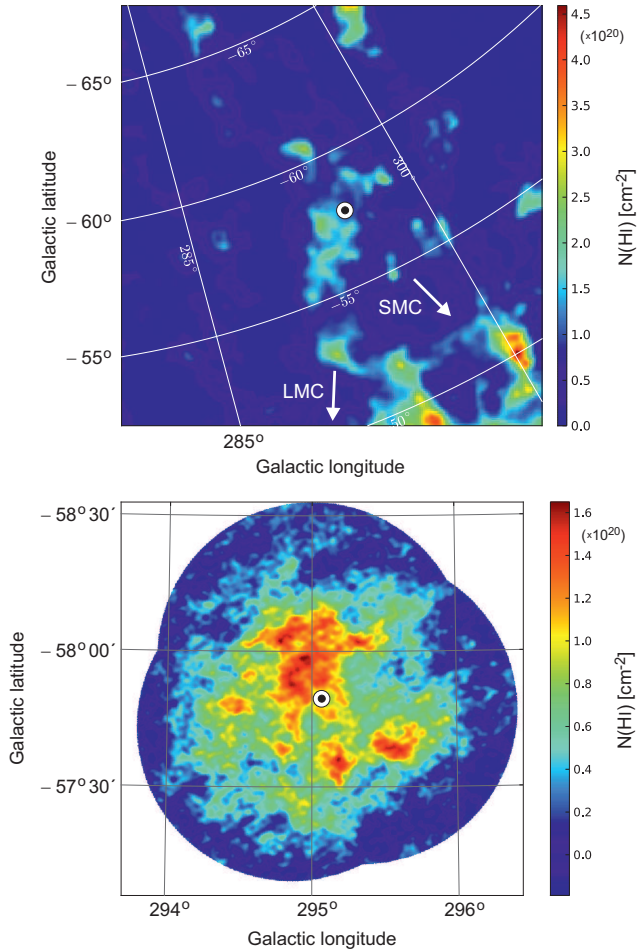}
\caption{{\it Upper panel:} H\,{\sc i} column-density map of the
Magellanic Stream in the general direction of Fairall\,9, based on
21cm data from GASS (the angular resolution is $15\farcm6$).
The map shows the distribution of neutral gas
in the LSR velocity range between $+100$ and $+250$ km\,s$^{-1}$.
The directions to LMC and SMC are indicated with the white arrows.
{\it Lower panel:} H\,{\sc i} column-density map of the
Fairall\,9 filament in the Stream, based on
21cm data from ATCA+Parkes (the angular resolution is
$1\farcm7 \times 1\farcm4$).
}
\end{figure}


\subsection{\emph{HST}/COS observations}

Fairall 9 was observed with the \emph{HST}/COS spectrograph 
(Green et al.\,2012) in 2012 under \emph{HST} program 
ID 12604 (PI: A. Fox). A four-orbit visit provided a total of
5378\,s of exposure time with the G130M/1291 wavelength setting, and
6144\,s with the G160M/1589 setting. With each grating,
all four FP-POS positions were used to dither the position of the
spectrum on the detector to reduce the fixed-pattern noise.
The raw data were processed and combined with the CALCOS 
pipeline (v2.17.3). For the coaddition of the individual
exposures we used interstellar absorption lines as 
wavelength reference. The final, co-added spectra then
were transformed into the LSR velocity frame. The COS data
cover the UV wavelength range between $1131$ and $1767$ \AA.
The spectra have a resolution of $R\approx16\,000$ 
(FWHM\,$\approx 19$ km\,s$^{-1}$) and a pixel size of $\sim 7$ 
km\,s$^{-1}$. The S/N ratio per resolution element is 
$37$ at $1200$ \AA\, and $26$ at $1550$ \AA.

In order to minimize geocoronal emission, which 
contaminates the absorption lines of 
O\,{\sc i} $\lambda 1302$, Si\,{\sc ii} $\lambda 1304$
and H\,{\sc i} Ly\,$\alpha$ in the velocity range
$-200$ km\,s$^{-1}\!\la\!v_{\rm LSR}\!\la\!+200$ km\,s$^{-1}$
during orbital daytime, we re-reduced the data with a night-only 
extraction. For this, data were extracted from those
time intervals when the Sun's altitude was less than $20$ degrees.

\subsection{\emph{FUSE} observations}

As part of our study we also re-analyze archival 
\emph{FUSE} spectra of Fairall\,9. These spectra were obtained in 2000 
under \emph{FUSE} program ID P101 (PI: K.R. Sembach) with a 
total exposure time of 34\,827\,s. They
show strong molecular hydrogen absorption arising in the MS,
as presented by Richter et al.\,(2001a).
The \emph{FUSE} spectra have a resolution of $\sim 20$ km\,s$^{-1}$ 
(FWHM), and cover the wavelength range $912-1180$ \AA. The raw data were 
reduced using the CALFUSE pipeline v3.2.1. The individual exposures 
were carefully coadded using interstellar lines as wavelength 
reference. Unfortunately, the S/N in the FUSE data is very low,
which severely hampers the analysis of absorption lines 
(see also Richter et al.\,2001a; Wakker 2006).
From a detailed inspection of the individual spectra
in the different segments we conclude that
only the data from the lithium-fluoride coated
segment 1A (LiF\,1A) can be properly coadded
without introducing large systematic errors in the flux
distribution in the spectrum. In the coadded LiF\,1A data, 
that are used by us to re-analyze the H$_2$ absorption in the MS,
we measure a S/N of $\sim 6$ per resolution element at $1020$ \AA.
Only for the wavelength range $1050-1082$ \AA\, does
the S/N ratio rise up to a maxium of $\sim 15$ per resolution
element due to the increased background flux from the broad, redshifted 
Ly\,$\beta$ emission from Fairall\,9.

\subsection{\emph{GASS} 21cm observations}

The H\,{\sc i} 21cm data for the Fairall\, sightline were taken from 
the Galactic-All Sky Survey (GASS, McClure-Griffiths et al.\,2009, 
Kalberla et al.\,2010). The survey was observed with the 64-m radio telescope at Parkes. 
The data cubes have an angular resolution of $15\farcm6$, leading to an 
RMS of $57\,\mathrm{mK}$ per spectral channel ($\Delta v=0.8\,\mathrm{km\,s}^{-1}$). 
This value translates to an H\,{\sc i} column density detection limit of 
$N$(H\,{\sc i}$)_{\rm lim}=4.1\times10^{18}$ cm$^{-2}$,
assuming a Gaussian-like emission line with a width of 
$20\,\mathrm{km\,s}^{-1}$ FWHM. Fig.\,1, upper panel, shows an H\,{\sc i} column
density map of the local environment of the Magellanic Stream 
centered on Fairall\,9 based on the GASS 21cm data. 

\subsection{ATCA 21cm data}

We supplement our measurements with higher-resolution H\,{\sc i} data
obtained with the Australia Compact Telescope Array (ATCA). The data in the
direction of Fairall\,9 were observed in 1998 and 1999 by Mary Putman
using the 750A, 750B, and 750D configurations. The ATCA is an east-west
interferometer with six antennas. Each antenna has a diameter of 22\,m.
For the observations a correlator band width of 4\,MHz was chosen,
resulting in a velocity resolution of about $0.8$ km\,s$^{-1}$.
Since the Fairall 9 cloud is too large to fit within a single pointing a
mosaic consisting of three fields was made. The observing time for each
of these three fields was 12\,hours leading to a full u-v coverage.
The FWHM of the synthesized ATCA beam is $1\farcm7 \times 1\farcm4$.

The ATCA data set was reduced by Christian Br\"uns with the MIRIAD
software. To circumvent the missing short-spacings problem, single-dish
data from the Parkes telescope were used. Image deconvolution and
combination with the single-dish data were performed with the
Miriad-Task MOSMEM. The Parkes data used for the short-spacings 
correction were obtained in the framework of an H\,{\sc i} survey of the 
Magellanic System (see Br\"uns et al.\,2005 for details).


\begin{figure}[t!]
\epsscale{0.90}
\plotone{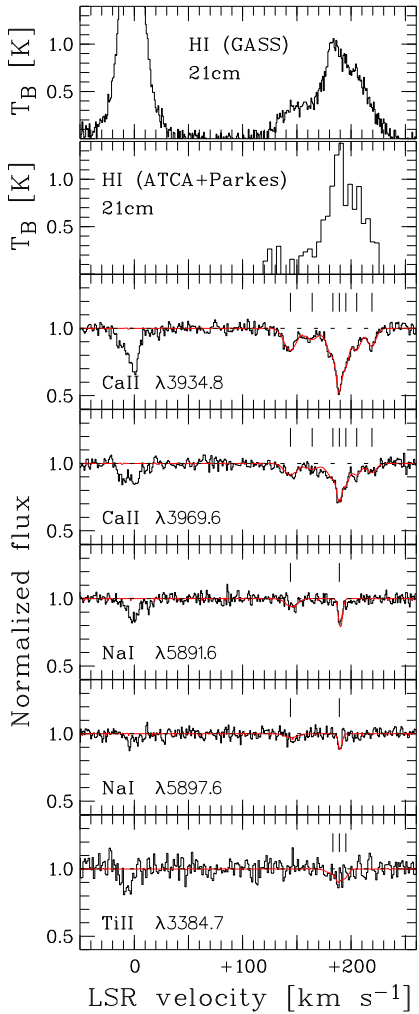}
\caption{Optical absorption profiles of Ca\,{\sc ii}, Na\,{\sc i},
and Ti\,{\sc ii} from VLT/UVES data of Fairall\,9 are shown.
Absorption from the Magellanic Stream is seen at LSR velocities between
$+130$ and $+240$ km\,s$^{-1}$. The solid red line display the
best possible Voigt profile fit to the data. The individual
velocity components are indicated by the tic marks. In the
upper two panels the H\,{\sc i} 21cm emission spectra toward
Fairall\,9 from GASS and ATCA data are plotted for comparison.
}
\end{figure}


\subsection{Spectral analysis methods}

Our strategy for the analysis of the optical and UV absorption-line
data of Fairall 9 combines different techniques to optimally account 
for the different spectral resolutions and S/N in the data.
The reduced and coadded spectra from UVES, COS, and \emph{FUSE} first were
continuum-normalized using low-order polynomials that were
were fit locally to the spectral regions of interest.

For the high-resolution VLT/UVES data we have used Voigt-profile 
fitting to decompose the MS absorption pattern in the optical lines
of Ca\,{\sc ii}, Na\,{\sc i}, and Ti\,{\sc ii} into 
individual absorption components (Voigt components) and to
derive column densities ($N$) and Doppler parameters ($b$ values) for
the individual components. For the fitting, we made use of 
the {\tt FITLYMAN} package implemented in the ESO-MIDAS analysis 
software (Fontana \& Ballester 1995). Laboratory wavelengths and 
oscillator strengths have been taken from the compilation
of Morton (2003).
Note that with our fitting technique we are able to measure $b$ values
smaller than the intrumental resolution as we {\it simultaneously}
fit the line doublets of Ca\,{\sc ii} and Na\,{\sc i}, so that
relative strengths of the lines are taken into account for the
determination of both $b$ and $N$. From this fitting procedure 
we obtain a component model for the Fairall\,9 sightline, in which 
the LSR velocity centroids of the individual absorption
components and the $b$ values for the low ions are defined.

In addition to Voigt-profile fitting, we have used the apparent
optical depth method (AOD method; Savage \& Sembach 1991) to derive
total gas column densities for the (unsaturated) optical absorption
profiles of Ca\,{\sc ii}, Na\,{\sc i}, and Ti\,{\sc ii}.
The AOD analysis was made using the custom-written
MIDAS code {\tt span} that allows us to measure equivalent widths
and AOD column densities (and their errors) in absorption spectra 
from a direct pixel integration. 

For the medium-resolution \emph{HST}/COS data we have used profile
{\it modeling} and the AOD method to derive column densities and 
column density limits for 
the various different low, intermediate, and high ions that have 
detectable transitions in the COS wavelength range.
The ion transitions in the COS data considered in this study
include C\,{\sc ii} $\lambda 1334.5$,
C\,{\sc ii}$^{\star}$ $\lambda 1335.7$,
C\,{\sc iv} $\lambda\lambda 1548.2,1550.8$,
N\,{\sc i}  $\lambda\lambda 1199.6,1200.2,1200.7$,
N\,{\sc v} $\lambda\lambda 1238.8,1242.8$,
O\,{\sc i} $\lambda 1302.2$,
Al\,{\sc ii} $\lambda 1670.8$,
Si\,{\sc ii} $\lambda\lambda 1190.4,1193.3,1260.4,1304.4,1526.7$,
Si\,{\sc iii} $\lambda 1206.5$,
Si\,{\sc iv} $\lambda\lambda 1393.8,1402.8$,
P\,{\sc ii} $\lambda 1152.8$,
S\,{\sc ii}  $\lambda\lambda 1250.6,1253.8,1259.5$,
Fe\,{\sc ii} $\lambda\lambda 1143.2,1144.9,1608.5$,
and Ni\,{\sc ii} $\lambda 1370.1$. For the profile modeling 
of neutral and singly-ionized species that trace predominantly
neutral gas in the MS we have used as input the component
model defined by the optical Ca\,{\sc ii} absorption (see above).
In this model, the LSR velocities and $b$ values
of the seven absorption components are constrained
by the Ca\,{\sc ii} fitting results, while the column
density for each component is the main free parameter 
that can be varied for each ion listed above. Our previous studies of 
Ca\,{\sc ii} in the Galactic halo (Richter et al.\,2005, 2009; 
Ben Bekhti et al.\,2008, 2011; Wakker et al.\,2007, 2008) 
and in intervening  absorbers at low redshift (Richter et al.\,2011)
have demonstrated that Ca\,{\sc ii} is an an excellent tracer 
for the distribution of neutral and partly ionized gas and 
its velocity-component structure, even in regions where 
Ca\,{\sc ii} is not the dominant ionization state.

Based on the Ca\,{\sc ii} model, and using a modified 
version of {\tt FITLYMAN}, we have calculated for each 
individual UV line a synthetic absorption profile, for 
which we have convolved an initial Voigt profile with 
input parameters $(v_i,b_i,N_i)$ for $i=1...7$ with 
the COS line-spread function (LSF) that is 
appropriate for the wavelength of the line. For 
the COS LSF we have used the improved LSF model
described by Kriss (2011). Then, the
columns $N_i$ were varied to minimize the differences 
between the synthetic absorption profile and the 
observed COS data. This method delivers reliable 
(total) column densities for those ions that have
multiple transitions with substantially 
different oscillator strengths in the COS wavelength 
range and for individual lines that are not fully 
saturated. More details about the accuracy of
this method are presented in Sect.\,3.1 and in
the Appendix.

For the intermediate and 
high ions (Si\,{\sc iii}, Si\,{\sc iv}, C\,{\sc iv})
the component structure and $b$ values are expected to be
different from that of the low ions (as the gas phase
traced by these ions often is spatially distinct from
the gas phase traced by the low ions), but no
information is available on the true component structure
of these ions from the optical data. Therefore,
we did not try to model the absorption profiles
of the high-ion lines, but estimated total column
densities (and limits) for these ions solely from the AOD 
method. Similarly, we used solely the AOD method to determine
the column density of C\,{\sc ii}$^{\star}$.

For the analysis of H$_2$ in the Magellanic 
Stream detected in the Fairall\,9 \emph{FUSE} data (Richter et al.\,2001a),
we have modeled the H$_2$ absorption using 
synthetic spectra generated with {\tt FITLYMAN}. As model
input we take into accout the component structure and line
widths seen in the optical Ca\,{\sc ii}/Na\,{\sc i} absorption, 
together with a Gaussian LSF according to the \emph{FUSE} spectral 
resolution. H$_2$ wavelengths and oscillator strengths 
have been adopted from Abgrall \& Roueff (1989).
Details on the H$_2$ modeling are presented 
in Sect.\,3.3.

%

\begin{figure}[t!]
\epsscale{1.2}
\plotone{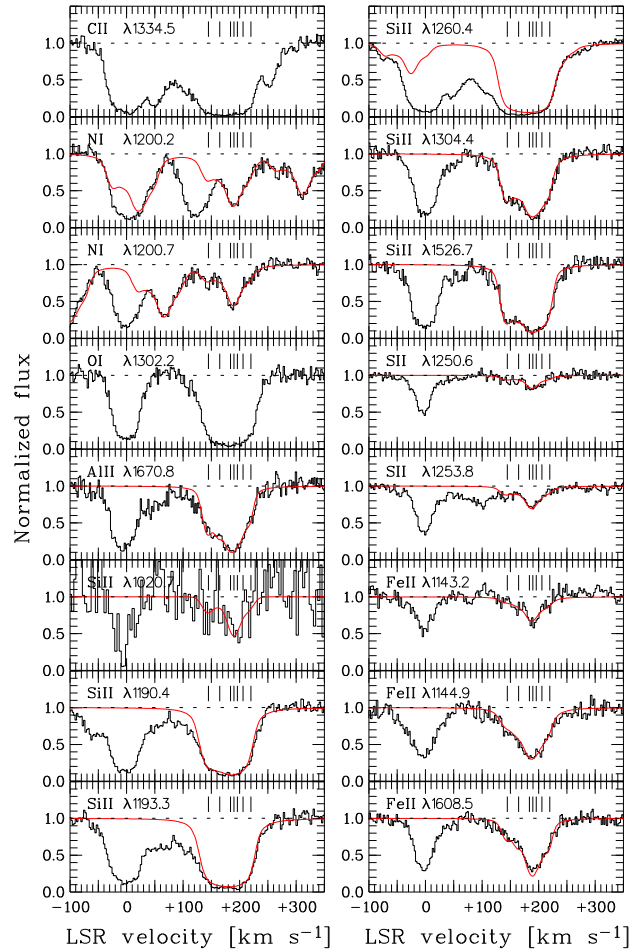}
\caption{Velocity profiles of various low ions in the COS
data of Fairall\,9 are shown. The FUV data for
Si\,{\sc ii} $\lambda 1020.7$ are from \emph{FUSE}. The velocity
components related to gas in the Magellanic Stream, as
identified in the optical Ca\,{\sc ii} data, are
indicated with the tic marks. The red solid line indicates
the best-fitting absorption model for the UV absorption, as
described in Sect.\,3.2. The single, saturated lines of
C\,{\sc ii} $\lambda 1334.5$ and O\,{\sc i} $\lambda 1302.2$
are not considered in our model.
}
\end{figure}


\vspace{0.6cm}

\section{Column-density measurements}

\subsection{Metal absorption in the optical}

Optical absorption related to gas in the Magellanic Stream
in the velocity range $120-230$ km\,s$^{-1}$ (this velocity
range is defined by the H\,{\sc i} 21\,cm data of the Stream) is
detected in the UVES spectrum of Fairall\,9 in the lines of
Ca\,{\sc ii} ($\lambda\lambda 3934.8,3969.6$), Na\,{\sc i}
($\lambda\lambda 5891.6,5897.6$), and Ti\,{\sc ii} ($\lambda 3384.7$).
Fig.\,2 shows the normalized absorption
profiles of these ions plotted on the LSR velocity scale together
with the H\,{\sc i} 21cm emission profiles from GASS and ATCA.
Ca\,{\sc ii} absorption is detected in seven individual
absorption components centered at $v_{\rm LSR}=+143,+163,+172,
+188, +194,+204$ and $+218$ km\,s$^{-1}$ with logarithmic 
column densities in the range log $N$(Ca\,{\sc ii}$)=11.29-11.68$
(where $N$ is in units [cm$^{-2}$] throughout the paper). The red solid
line in Fig.\,2 indicates the best Voigt-profile fit to
the data. All Ca\,{\sc ii}, Na\,{\sc i} and Ti\,{\sc ii}
column-density measurements are summarized in Table 1.

%

\begin{deluxetable*}{llrrrrrrrrr}
\tabletypesize{\small}
\tablewidth{0pt}
\tablecaption{Summary of ion column-density measurements}
\tablehead{
\colhead{} &
\colhead{Component} & \colhead{1} & \colhead{2} & \colhead{3} & \colhead{4} & \colhead{5} &
\colhead{6} & \colhead{7}\\
\colhead{} &
\colhead{$v$ [km\,s$^{-1}$]} & \colhead{$+143$} & \colhead{$+163$} & \colhead{$+172$} & \colhead{$+188$} &
\colhead{$+194$} & \colhead{$+204$} & \colhead{$+218$} & & \\
\colhead{} &
\colhead{$b$ [km\,s$^{-1}$]} & \colhead{$7.4$} & \colhead{$7.9$} & \colhead{$6.4$} & \colhead{$1.8$} &
\colhead{$3.4$} & \colhead{$4.9$} & \colhead{$5.1$} & & \\
\colhead{} &
\colhead{$\sigma_b$ [km\,s$^{-1}$]} & \colhead{$2.4$} & \colhead{$4.1$} & \colhead{$2.2$} & \colhead{$0.3$} &
\colhead{$1.4$} & \colhead{$1.7$} & \colhead{$2.0$} & & \\
\\
\colhead{Ion} &
\colhead{Instrument} & \colhead{log\,$N_1$} & \colhead{log\,$N_2$} & \colhead{log\,$N_3$} & \colhead{log\,$N_4$} &
\colhead{log\,$N_5$} & \colhead{log\,$N_6$} & \colhead{log\,$N_7$} & \colhead{log\,$N_{\rm tot,fit}$} & 
\colhead{log\,$N_{\rm tot,AOD}$}
}
\startdata

H\,{\sc i}    & GASS     &  ... & ... & ... & ... & ... & ... & ... &
              ... & $19.95$ \\
 
\\
Na\,{\sc i}   & VLT/UVES & $10.92$ & ... & ... & $11.05$ & ... & ... & ... &
              $11.29\pm0.04$ & $11.33\pm0.05$ \\
Ca\,{\sc ii}  & VLT/UVES & $11.58$ & $11.29$ & $11.63$ & $11.68$ & $11.58$ & $11.43$ & $11.35$ &
              $12.37\pm0.02$ & $12.38\pm0.03$ \\
Ti\,{\sc ii}  & VLT/UVES & ... & ... & $11.11$ & $11.15$ & $11.24$ & ... & ... &
              $11.27\pm0.06$ & $\leq 12.00$ \\

\\
C\,{\sc i}    & \emph{HST}/COS & ... & ... & ... & ... & ... & ... & ... &
              ... & $\leq 13.02$ \\
C\,{\sc ii}   & \emph{HST}/COS & ... & ... & ... & ... & ... & ... & ... &
              ... & $\geq 14.93$ \\
C\,{\sc ii}$^{\star}$ & \emph{HST}/COS & ... & ... & ... & ... & ... & ... & ... &
              ... & $13.35\pm0.07$ \\

C\,{\sc iv}   & \emph{HST}/COS & ... & ... & ... & ... & ... & ... & ... &
              ... & $13.73\pm0.04$ \\
N\,{\sc i}    & \emph{HST}/COS & $13.65$ & $13.42$ & $13.84$ & $14.10$ & $13.98$ & $13.54$ & $13.46$ &
              $14.63\pm0.05$ & $\geq 14.51$ \\
O\,{\sc i}    & \emph{HST}/COS & ... & ... & ... & ... & ... & ... & ... &
              ... & $\geq 14.77$ \\
Al\,{\sc ii}  & \emph{HST}/COS & $12.55$ & $12.53$ & $12.92$ & $12.58$ & $12.69$ & $12.37$ & $12.24$ &
              $13.45\pm0.06$ & $\geq 13.18$ \\
Si\,{\sc ii}  & \emph{HST}/COS & $14.10$ & $13.75$ & $14.17$ & $14.13$ & $14.24$ & $13.92$ & $13.89$ &
              $14.90\pm0.04$ & $15.01\pm0.17$ \\
Si\,{\sc iii} & \emph{HST}/COS & ... & ... & ... & ... & ... & ... & ... &
              ... & $\geq 13.57$ \\
Si\,{\sc iv}  & \emph{HST}/COS & ... & ... & ... & ... & ... & ... & ... &
              ... & $13.03\pm0.06$ \\
P\,{\sc ii}   & \emph{HST}/COS & ... & ... & ... & ... & ... & ... & ... &
              ... & $\leq 12.87$ \\
S\,{\sc ii}   & \emph{HST}/COS & $13.95$ & $13.70$ & $14.03$ & $14.00$ & $14.09$ & $13.77$ & $13.74$ &
              $14.77\pm0.02$ & $14.77\pm0.04$ \\
Fe\,{\sc ii}  & \emph{HST}/COS & $13.47$ & $13.61$ & $13.94$ & $13.85$ & $13.94$ & $13.67$ & $13.65$ &
              $14.61\pm0.04$ & $14.59\pm0.04$ \\
Ni\,{\sc ii}  & \emph{HST}/COS & ... & ... & ... & ... & ... & ... & ... &
              ... & $\leq 13.37$ \\
\enddata
\end{deluxetable*}

%

The strongest Ca\,{\sc ii}
absorption component in the MS (in terms of the absorption depth)
is component 4 at $+188$ km\,s$^{-1}$; this very narrow
component is also detected in both Na\,{\sc i} lines (Fig.\,2).
From the simultaneous fit of the Ca\,{\sc ii} and Na\,{\sc i} doublets
we obtain a very small Doppler parameter for component 4 of
$b=1.8$ km\,s$^{-1}$. The  small $b$ value for this component
is confirmed by fits to the individual lines of Ca\,{\sc ii} 
and Na\,{\sc i}, which all imply $b<2$ km\,s$^{-1}$.
The detection of Na\,{\sc i} together with the
small $b$ value indicates that the gas in component 4 is
relatively cold and dense and possibly is confined in a
dense core with little turbulence. 
Components 3,4, and 5 are also detected 
in Ti\,{\sc ii} (Fig.\,2).
Since Ti\,{\sc ii} and H\,{\sc i} have almost identical ionization
potentials (Table 2), the detection of Ti\,{\sc ii} 
suggests that most of the neutral
gas column density is contained in these three components.
This conclusion is supported by the H\,{\sc i} 21cm emission profiles
from GASS and ATCA (Fig.\,2; first two panels), which also have their maxima
in the velocity range between $+170$ and $+200$ km\,s$^{-1}$.
Weak and relatively broad ($b=7.4$ km\,s$^{-1}$) Na\,{\sc i} absorption
is also detected in component 1 at $+143$ km\,s$^{-1}$.

As mentioned in Sect.\,2.5, we also have used the AOD method to determine
the total column densities for Ca\,{\sc ii}, Na\,{\sc i} and Ti\,{\sc ii}
by integrating over the velocity range relevant for MS absorption
(i.e., $+130$ km\,s$^{-1} \leq v_{\rm LSR} \leq +230$ km\,s$^{-1}$.)
The column densities derived by these two different methods agree
very well within their $1 \sigma$ error ranges (see Table 1, rows 10 and 11).

\subsection{Metal absorption in the UV}

In Fig.\,3 we show normalized UV absorption profiles of the low 
ions C\,{\sc ii}, N\,{\sc i}, O\,{\sc i}, Al\,{\sc ii}, Si\,{\sc ii}, 
S\,{\sc ii}, and Fe\,{\sc ii}. The absorption profiles of the 
intermediate and high ions Si\,{\sc iii}, C\,{\sc iv} 
and Si\,{\sc iv} are shown in Fig.\,4.
Following the procedure described in Sect.\,2.5, we have reconstructed the
absorption pattern of UV metal lines of the low ions,
using the component model defined by the optical lines of Ca\,{\sc ii}. 
As already mentioned, the modeling method provides relevant results
only for those ions that have multiple transitions in the available 
COS wavelength band and for single lines that are not (or at most mildly)
saturated. In our case, the modeling method could be used to determine
column densities for N\,{\sc i}, Al\,{\sc ii}, Si\,{\sc ii}, S\,{\sc ii},
and Fe\,{\sc ii}, while for the fully saturated (single) lines of 
C\,{\sc ii} and O\,{\sc i} the modeling does not yield relevant 
column density limits.
In Fig.\,3, the best-fitting column-density model for each ion is overlaid 
with the red solid line. The best-fitting model assumes the same $b$ values
for the individual subcomponents as derived for Ca\,{\sc ii} from
the Voigt fit of the UVES data. The ion column densities for each velocity
component are listed in Table\,1 in rows $2-7$; the total column density
for each ion is given in row\,10. As can be seen in Fig.\,3, 
the absorber model successfully reproduces the shape of the absorption 
lines in the UV. 

From Table\,1 follows that the relative column-densities in the
seven velocity components differ slightly from ion to ion. These
differences are expected, because most ions have different ionization
potentials and therefore trace slightly different gas phases within the
absorber. In addition, differential dust depletion for some elements
may also be relevant in this context.
Because of the relatively limited spectral resolution of the COS instrument 
and the resulting overlap between  neighbouring velocity components 
the accuracy for the determination of the column densities for individual 
subcomponents is relatively low ($\sim 0.30$ dex, typically).
Much better constrained are the {\it total} column densities (Table 1, row 10)
for the ion multiplets N\,{\sc i}, S\,{\sc ii}, Si\,{\sc ii} and Fe\,{\sc ii},
which have $1\sigma$ uncertainties $<0.05$ dex.
This is because the integrated (total) column density for each of these 
ions is determined predominantly by the integrated equivalent widths 
of the weaker lines and the equivalent width {\it ratios}
of the weaker and stronger transitions.
In the Appendix we present additonal model plots to provide a more 
detailed insight into the allowed parameter range $(b_i,N_i)$
for the individual subcomponents and their impact 
on the shape of the absorption profiles and total column-density
estimates.

Next to the absorption modeling we have used the AOD method to determine
total column densities (or column-density limits) in the COS data for each of
the ions listed above (Table\,1, row\,11). For the unsaturated lines of
S\,{\sc ii}, Si\,{\sc ii}, and Fe\,{\sc ii} the total column densities derived
from the AOD method are in excellent agreement ($< 1\sigma$)
with the total column densities determined from the profile modeling.
Note that the absorption of C\,{\sc ii}$^{\star}$ in the MS will
be discussed separately in Sect.\,4.4.

%

\begin{figure}[t!]
\epsscale{1.0}
\plotone{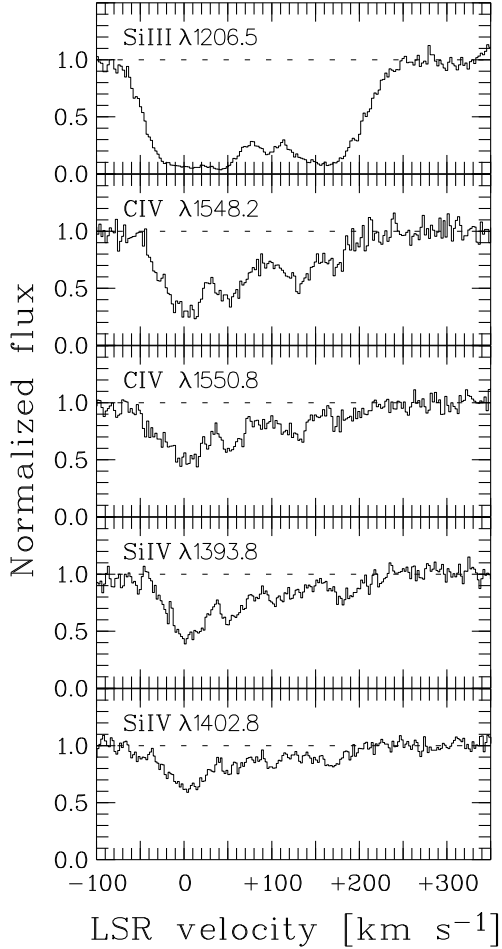}
\caption{COS velocity profiles of the intermediate and
high ions Si\,{\sc iii}, C\,{\sc iv} and Si\,{\sc iv}. The
velocity structure in these ions is very different from that
of the low ions, suggesting that they trace spatially
distinct regions within the MS. High-velocity Si\,{\sc iv}
absorption between $+150$ and $+200$ km\,s$^{-1}$ in the
Si\,{\sc iv} $\lambda 1402.8$ line is possibly blended with
intergalactic absorption.}
\end{figure}

%

\subsection{H\,{\sc i} 21cm emission}

The GASS 21cm velocity profile shown in the upper panel of Fig.\,2
indicates H\,{\sc i} emission from gas in the Magellanic Stream
in the LSR velocity range between $+100$ and $+250$ km\,s$^{-1}$. 
The MS emission shows an asymmetric pattern with a peak
near $v_{\rm LSR}\approx +180$ km\,s$^{-1}$, surrounded
by weaker satellite componets at lower and higher radial 
velocities. The overall shape of the emission profile
mimics that of the inversed absorption profiles of
the low-ion lines of N\,{\sc i} and S\,{\sc ii}
indicating that both the pencil-beam absorption data and 
the GASS emission data, that have a beam-size of $15\farcm6$,
sample the same physical structure. Integrating the GASS data 
in the above given velocity range for the single pixel that
covers the position of Fairall\,9 we obain a total H\,{\sc i}
column density in the Stream of log $N$(H\,{\sc i}$)=19.95$
($N$(H\,{\sc i}$)=(8.96\pm0.09)\times 10^{19}$ cm$^{-2}$).
The pixel value is somewhat smaller than the beam-averaged 
column density, which is log $N$(H\,{\sc i}$)=19.98$.

In the second panel of Fig.\,2 we show the 21cm emission
profile for the Fairall\,9 direction from the high-resolution ATCA data  
in the LSR velocity range between $+120$ and $+230$ km\,s$^{-1}$.
Because of the relatively low S/N in the ATCA data, the 
spectrum shown in Fig.\,2 was binned for displaying purposes 
to $8.3$ km\,s$^{-1}$ wide pixels.
The overall shape of the profile from the ATCA data 
is roughly similar to the one from GASS, but in the ATCA 
spectrum the MS 21cm emission peaks 
at a somewhat higher velocity ($v_{\rm LSR}\approx +190$ km\,s$^{-1}$, 
thus coinciding with the peak absorption in Ca\,{\sc ii} and
Na\,{\sc i}; see Fig.\,2) and at a slightly higher 
brightness temperature. The MS emission profile from ATCA 
is, however, somewhat narrower than the GASS profile. Integration
over the MS velocity range in the (unbinned) ATCA data yields a 
total H\,{\sc i} column density in the Stream of 
$N$(H\,{\sc i}$)=(8.6\pm1.5)\times 10^{19}$ cm$^{-2}$ or
log $N$(H\,{\sc i}$)=19.93^{+0.07}_{-0.08}$, thus in
excellent agreement with the GASS data.

For comparison, Gibson et al.\,(2000) find 
log $N$(H\,{\sc i}$)=19.97$ 
($N$(H\,{\sc i}$)=(9.35\pm0.47)\times 10^{19}$ cm$^{-2}$)
using Parkes 21cm observations of Fairall\,9 with a 
beam size of $14\farcm0$. From the 
21cm data of the {\it Leiden-Argentine-Bonn Survey} (LAB;
Kalberla et al.\,2005) with $\sim 0.5\deg$ resolution we 
obtain for the same velocity range an H\,{\sc i} 
column density in the Stream of log $N$(H\,{\sc i}$)=19.97$.

The H\,{\sc i} column densities in the MS toward Fairall\,9
derived from different instruments with very different
angular resolutions agree with each other
within their $1\sigma$ error ranges. Our conclusion is that 
beam smearing is not expected to be 
a critical issue for the determination of metal
abundances in the MS toward Fairall\,9 
(using 21cm data in combination with the pencil-beam
UVES and COS data). In the following, we adopt the value of 
log $N$(H\,{\sc i}$)=19.95$ for the total H\,{\sc i} column 
density in the Stream toward Fairall\,9 together
with an appropriate estimate of the systematic error
of $0.03$ dex that accounts for the limited spatial
resolution of the 21cm data. 

As it is shown in 
the Appendix, this H\,{\sc i} column density in the MS
is in agreement with the observed shape of the H\,{\sc i}
Ly\,$\alpha$ absorption line in the COS data of Fairall\,9, 
which provides an independent (albeit less stringent) measure 
for the neutral-gas column density in the Stream in 
this direction (see also Wakker, Lockman, \& Brown 2011
for a detailed discussion on this topic).

\subsection{Molecular hydrogen absorption}

The detection of H$_2$ absorption in the Magellanic Stream in the 
\emph{FUSE} data of Fairall\,9 has been reported previously by 
Richter et al.\,(2001a) and Wakker (2006). 
H$_2$ absorption is seen near $v_{\rm LSR}=+190$ km\,s$^{-1}$ in the
rotational levels $J=0,1$ and $2$. Although the overall S/N in
the \emph{FUSE} data is low ($\leq15$ per resolution element;
see Sect.\,2.5), the H$_2$ absorption
is clearly visible in many lines due to the relatively high
H$_2$ column density (Richter et al.\,2001a). 

%

\begin{figure*}[t!]
\epsscale{1.0}
\plotone{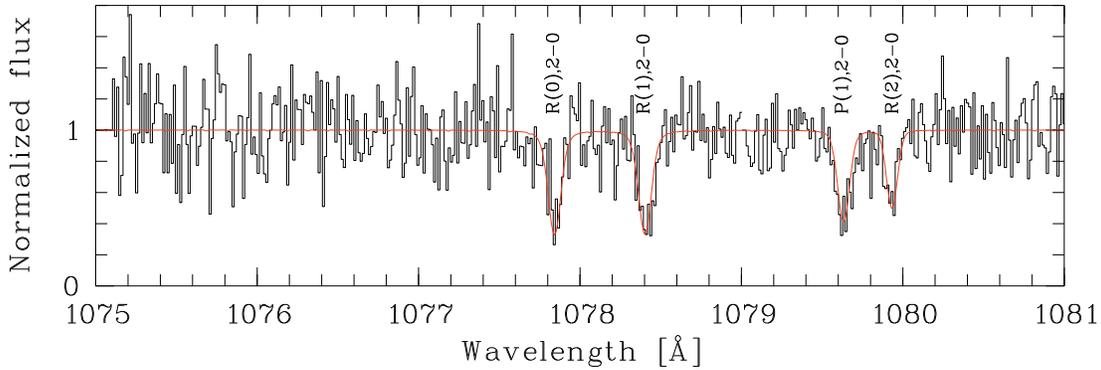}
\caption{Continuum-normalized \emph{FUSE} spectrum of Fairall\,9 in the wavelength
range between $1075$ and $1081$ \AA. Absorption by H$_2$ from the rotational
levels $J=0,1$ and $2$ related to gas in the Magellanic Stream are indicated
above the spectrum using the common notation scheme. The best-fitting
model for the H$_2$ absorption is overplotted with the red solid line.
}
\end{figure*}

%

%

\begin{deluxetable}{lrrl}
\tablewidth{0pt}
\tablecaption{Summary of H$_2$ column-density modeling}
\tablehead{
\colhead{$J$} & $b$$^{\rm a}$ [km\,s$^{-1}$] & 
\colhead{log $N(J)$$^{\rm b}$} & \colhead{$T_J$ [K]$^{\rm c}$} \\
}
\startdata
$0$ & $2.0$ & $17.53\pm 0.24$ & $93^{+149}_{-39}$ \\
$1$ & $2.0$ & $17.68\pm 0.26$ & $93^{+149}_{-39}$ \\
$2$ & $2.0$ & $16.46\pm 0.21$ & $<172$ \\
$3$ & $2.0$ & $\leq 15.60$    & $<172$ \\
Sum &       & $17.93^{+0.19}_{-0.16}$\\
\enddata
\tablenotetext{a}{$b$-value fixed to $2.0$ km\,s$^{-1}$; see Sect.\,3.4.}
\tablenotetext{b}{Logarithmic H$_2$ column density for rotational state $J$.}
\tablenotetext{c}{Equivalent Boltzmann temperature implied by $N(J)$; see Sect.\,4.4.2.}
\end{deluxetable}

%

We have reanalyzed the Fairall\,9 \emph{FUSE} data based on an improved
data reduction pipeline and the component model
discussed in the previous sections. Richter et al.\,(2001a)
have used a curve-of-growth technique to derive the column densities
$N(J)$ and $b$ values of the H$_2$ rotational states $J=0,1,2$, assuming
as single MS absorption component. In contrast, we here model the
absorption spectrum of the MS H$_2$ absorption in the \emph{FUSE} Fairall\,9
data using synthetic spectra. For our model we assume
that the H$_2$ absorption arises in the region with the highest
gas density in the Stream, which is component 4 at $v_{\rm LSR}=+188$
km\,s$^{-1}$ detected in the UVES data in 
Ca\,{\sc ii}, Na\,{\sc i}, and Ti\,{\sc ii} (Table\,1).
In fact, from absorption studies of
Na\,{\sc i} and H$_2$ in interstellar gas it is well known that these
two species trace the same gas phase (i.e., cold neutral and cold
molecular gas; e.g., Welty et al.\,2006).

We adopt $b_{\rm turb}=1.8$ km\,s$^{-1}$ for the H$_2$ absorption
in component 4, as suggested by the Na\,{\sc i} and Ca\,{\sc ii}
absorption in the same component (Table\,1). We further assume
(and prove later) that the kinetic temperature of the H$_2$
absorbing gas is $T_{\rm kin}\approx100$ K in component 4. 
It is usually assumed that the Doppler parameter is composed 
of a turbulent and a thermal component, so that
$b^2=b_{\rm turb}^2+b_{\rm th}^2$. For light elements
and molecules (such as H$_2$) $b_{\rm th}$ must not be 
neglected. We therefore adopt $b$(H$_2)=2.0$\,km\,s$^{-1}$ 
for our H$_2$ model spectrum.

Fig.\,5 shows our best fitting H$_2$ model (red solid line)
together with the \emph{FUSE} data for the wavelength range between
$1070$ and $1080$ \AA. Since the $b$ value is fixed to
$b$(H$_2)=2.0$\,km\,s$^{-1}$ and the velocity to 
$v_{\rm LSR}=+188$ km\,s$^{-1}$, the only free parameters in
our model are the H$_2$ column densities $N(J)$. Table\,2
summarizes the best fitting H$_2$ column densities using
this method. The given errors represent the 
uncertainties for log $N(J)$ under the assumption
that $b=2.0$ km\,s$^{-1}$. The resulting total H$_2$ column
density is log $N$(H$_2)=17.93^{+0.19}_{-0.16}$ and 
the fraction of hydrogen in molecular form is
log $f_{\rm H_2}=$\,log\,$[2N($H$_2)/(N$(H\,{\sc i}$)+2N($H$_2))]=-2.03$.
Note that the value for log $f_{\rm H_2}$ derived in this way 
represents a sightline-average; the local molecular fraction
in the H$_2$ absorbing region must be higher
(because $N$(H\,{\sc i}) in this region is smaller).

The H$_2$ column densities we derive from our spectral modeling
are substantially ($\sim 1$ dex) higher than the values presented
in the earlier studies of the Fairall\,9 \emph{FUSE} data
(Richter et al.\,2001a; Wakker 2006). The
reason for this discrepancy is the substantially lower
$b$ value that we adopt here. This more realistic $b$ value 
for the H$_2$ absorbing gas is a direct consequence of the 
previously unknown complex absorption pattern in the MS 
toward Fairall\,9, that is visible only in the high-resolution optical 
data of Ca\,{\sc ii} and Na\,{\sc i} (see Sect.\,3.1). Our study
suggests, that, without the support of high-resolution and
high S/N optical or UV data that can be used to resolve
the true sub-component structure of the absorbers and to deliver
reliable $b$ values for the subcomponents, the interpretation of
interstellar H$_2$ absorption lines in low-resolution and low-S/N data
from \emph{FUSE} can be afflicted with large systematic uncertainties.

%

\section{Physical and chemical conditions in the gas}

\subsection{Ionization conditions and overall metallicity}

The presence of high, intermediate and low ions, as well as 
molecular hydrogen, in the Magellanic Stream toward Fairall\,9
in seven individual velocity components implies a complex 
multi-phase nature of the gas. To obtain meaningful results
for the chemical composition and physical conditions 
from the measured ion column densities it is necessary to 
consider in detail the ionization conditions in the 
absorbing gas structure. For an element M we define
its relative gas-phase abundance compared to the 
solar abundance in the standard way
[M/H]\,=\,log\,(M/H)\,$-$\,log\,(M/H)$_{\sun}$.
Solar reference abundances have been adopted from
Asplund et al.\,(2009).

The best ion to study the overall metallicity of neutral and weakly 
ionized interstellar gas is O\,{\sc i}.
Neutral oxygen and neutral hydrogen have similar ionization
potentials and there is a strong charge-exchange reaction that
couples both ions in gas with sufficiently high density. 
For our sightline, however, the O\,{\sc i} column density 
is not well constrained because of the saturation of the 
O\,{\sc i} $\lambda 1302.2$ line and the lack of other,
weaker O\,{\sc i} transitions in the COS wavelength range.
Because it is impossible to measure the O\,{\sc i} column density 
in the MS using our component models and the AOD method to 
a satisfactory accuracy, we do not further consider the 
O\,{\sc i} $\lambda 1302.2$ line for our analysis.

For the determination of the overall metallicity of the gas in
the Magellanic Stream toward Fairall\,9 we choose instead
S\,{\sc ii}, which is another useful ion for measuring interstellar
abundances, as singly-ionized sulfur is an excellent tracer for
neutral hydrogen without being depleted into dust grains
(e.g., Savage \& Sembach 1996). We also
have a very accurate column-density determination 
of log $N$(S\,{\sc ii}$)=14.77\pm 0.02$ in the Magellanic Stream,
based on the absorption modeling of the two S\,{\sc ii} lines
at $1250.6$ and $1253.8$ \AA\, and (independently) from
the AOD method (see Table 1). 
Our value is slightly higher than the S\,{\sc ii} column density
of log $N$(S\,{\sc ii}$)=14.69\pm 0.03$ derived by Gibson et al.\,(2000)
from \emph{HST}/GHRS data.

From the measured (S\,{\sc ii}/H\,{\sc i}) ratio we obtain an initial
estimate of the sulfur abundance in the Stream toward Fairall\,9 
of [S/H$]\approx -0.30$. Since the ionization 
potential of S\,{\sc ii} (23.2 eV; see Table 3) is higher
than that of H\,{\sc i} (13.6 eV), we have to model the ionization
conditions in the gas to obtain a more precise estimate for 
the sulfur abundance (and the abundances of the other detected 
metals) in the Stream toward Fairall\,9. While previous studies
of Galactic halo clouds (Wakker et al.\,1999; Gibson et al.\,2000) 
have demonstrated that the (S\,{\sc ii}/H\,{\sc i}) ratio is
a very robust measure for the sulfur-to-hydrogen ratio,
hardly being affected by the local ionizing radiation field,
the ionization modeling is important to obtain a relibale estimate
of the systematic uncertainty that arises from the use of 
S\,{\sc ii} as reference indicator for the overall $\alpha$
abundance in the gas. For this task, we use the 
photoionization code Cloudy (v10.00; Ferland et al.\,1998),
which calculates the expected column densities for different ions
as a function of the ionization parameter $U=n_{\gamma}/n_{\rm H}$ 
for a gas slab with a given neutral gas column density
and metallicity, assuming that the slab is illuminated 
by an external radiation field. As radiation field we adopt 
a combined Milky Way plus extragalactic ionizing 
radiation field based on the models by Fox et al.\,(2005) 
and Bland-Hawthorn \& Maloney (1999,2002),
appropriately adjusted to the position of the MS relative
to the Milky Way disk.

%

\begin{figure}[t!]
\epsscale{1.2}
\plotone{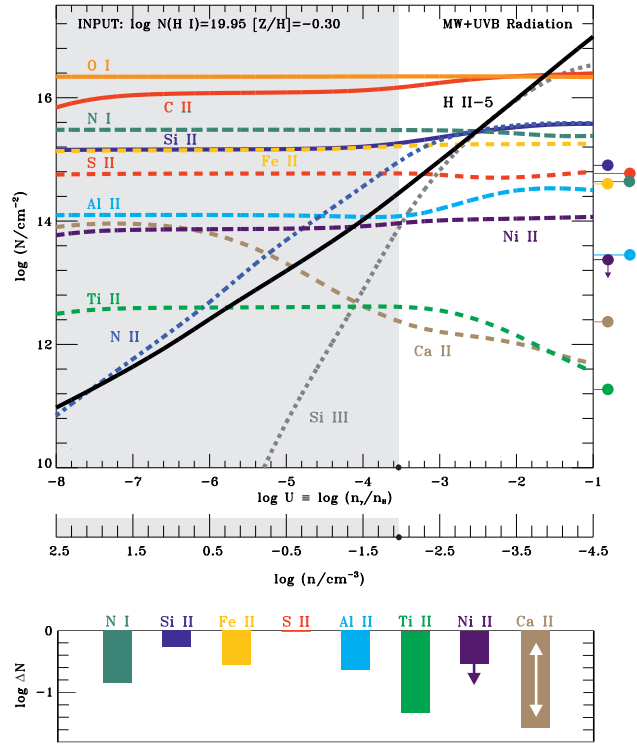}
\caption{
{\it Upper panel:} Cloudy photoionization model of low-ionization species
in the MS toward Fairall\,9. The colored lines indicate the predicted
ion column densities as a function of the ionization parameter $U$ and
gas density $n_{\rm H}$ for a gas slab with a (total) neutral hydrogen
column density of log $N$(H\,{\sc i}$)=19.95$ and a metallicity
of [Z/H$]=-0.30$. Observed ion column densities are
indicated on the right-hand side with the
filled colored circles. The gray-shaded area marks the
range for $U$ and $n_{\rm H}$ that is relevant for our study
based on the oberved Ca\,{\sc ii} column density (see Sect.\,4.1).
{\it Lower panel:} differences between the observed and the predicted
ion column densities, $\Delta N=N_{\rm Cloudy} - N_{\rm obs.}$; they
are discussed in Sect.\,4.1. For Ni\,{\sc ii}, the bar and the arrow indicate an
upper limit for $\Delta N$(Ni\,{\sc ii}), while for Ca\,{\sc ii} the bar
and the double-arrow indicate the allowed range for $\Delta N$(Ca\,{\sc ii}).
}

\end{figure}

%

While our absorption modeling provides rough estimates for the
ion column densities in the seven sub-components, we refrain
from trying to model the ionization conditions in the 
individual components. This is because the unknown H\,{\sc i}
column densities in these components, together with the unknown 
geometry of the overall gas structure, would lead to large
systematic uncertainties for such models.
Instead, we have used Cloudy to obtain {\it integrated} 
elemental abundances in the gas of the Magellanic Stream 
toward Fairall\,9 (representing the optical-depth weighted 
mean of the individual element abundances in all subcomponents).
It is, however, important to emphasize that meaningful results for 
integrated elemental abundances from Cloudy can be obtained only 
for those ions for which the column densities in the ionization
model do not strongly depend on the gas density (and ionization parameter), 
because the latter quantities are expected to vary {\it substantially} 
among the individual subcomponents. 

For our Cloudy model we assume a neutral gas column density
of log $N$(H\,{\sc i}$)=19.95$, an overall metallicity of 
[Z/H$]=-0.3$, and solar {\it relative} abundances of the
metals from Asplund et al.\,(2009), based on our results described above.
In Fig.\,6, upper panel, we display the expected logarithmic column
densities for low and intermediate ions as a function of log $U$
and log $n_{\rm H}$ for this model. The measured total column densities 
for these ions (Table 1, 10th column) are plotted on the right hand side
of the panel. The total Ca\,{\sc ii} column density of 
log $N$(Ca\,{\sc ii}$)=12.37$ sets a limit for the
(averaged) ionisation parameter and density in the gas 
of log $\langle U \rangle \leq -3.55$ and log $\langle n_{\rm H} \rangle
\geq -1.95$. The corresponding density range
that we consider as relevant for our Cloudy model 
($2.5\leq$\,log $n_{\rm H}$\,$\leq -1.95$) is indicated 
in Fig.\,6 with the gray-shaded area.
For this range, the expected column densities of N\,{\sc i}, 
Si\,{\sc ii}, Fe\,{\sc ii}, S\,{\sc ii}, Al\,{\sc ii}, 
Ni\,{\sc ii},  and Ti\,{\sc ii} are nearly
independent of log $U$ and log $n_{\rm H}$. For S\,{\sc ii},
in particular, the column density varies within only $0.03$ dex
in the above given density range, supporting the previous
conclusions from Wakker et al.\,(1999) and Gibson et al.\,(2000).
In the lower panel of Fig.\,6 we show the differences ($\Delta N$)
between the measured column densities and the mean column 
densities predicted by Cloudy. For Ni\,{\sc ii}
and Ca\,{\sc ii}, the arrows indicate the upper limit and the 
allowed range for $\Delta N$, respectively.

%

\begin{deluxetable*}{lrrrlll}
\tablewidth{0pt}
\tablecaption{Summary of integrated gas-phase abundances and dust depletion values}
\tablehead{
\colhead{Ion (X)} & \colhead{$Z$} & \colhead{I.P.$^{\rm a}$} & 
\colhead{log\,(M/H)$_{\sun}^{\rm b}$} & 
\colhead{log\,(X/H\,{\sc i})$^{\rm c}$} & \colhead{[M/H]$^{\rm d}$} &
\colhead{log $\delta$(M)$^{\rm e}$} \\
\colhead{} & \colhead{} & \colhead{[eV]} & \colhead{$+12$} &
\colhead{} & \colhead{} & \colhead{} \\
}
\startdata
C\,{\sc i}   &  6 & 11.26 & $8.43\pm0.05$ & $\leq -6.93$   & ... & ... \\
C\,{\sc ii}  &  6 & 24.38 & $8.43\pm0.05$ & $\geq -5.02$   & ... & ...     \\
N\,{\sc i}   &  7 & 14.53 & $7.83\pm0.05$ & $-5.32\pm0.05$ & $-1.15\pm0.06$ & ...\\
O\,{\sc i}   &  8 & 13.62 & $8.69\pm0.05$ & $\geq -1.87$   & ... & ...    \\
Na\,{\sc i}  & 11 &  5.14 & $6.24\pm0.04$ & $-8.66\pm0.04$ & ... & ...    \\
Al\,{\sc ii} & 13 & 18.83 & $6.45\pm0.03$ & $-6.50\pm0.06$ & $-0.92\pm0.08$ & $-0.62$ \\
Si\,{\sc ii} & 14 & 16.35 & $7.51\pm0.03$ & $-5.05\pm0.04$ & $-0.57\pm0.06$ & $-0.27$ \\
P\,{\sc ii}  & 15 & 19.77 & $5.41\pm0.04$ & $\leq -7.08$   & ... & ...    \\  
S\,{\sc ii}  & 16 & 23.34 & $7.12\pm0.03$ & $-5.18\pm0.02$ & $-0.30\pm0.04$ & $(0)$ \\
Ca\,{\sc ii} & 20 & 11.87 & $6.34\pm0.04$ & $-7.58\pm0.02$ & $\leq -0.30$ & $0$ to $-1.58$ \\
Ti\,{\sc ii} & 22 & 13.58 & $4.95\pm0.05$ & $-8.68\pm0.06$ & $-1.63\pm0.07$ & $-1.33$ \\
Fe\,{\sc ii} & 26 & 16.12 & $7.50\pm0.04$ & $-5.34\pm0.04$ & $-0.86\pm0.06$ & $-0.56$ \\
Ni\,{\sc ii} & 28 & 18.17 & $6.22\pm0.04$ & $\leq -6.58$   & $\leq -0.81$ & $\leq -0.51$ \\
\enddata
\tablenotetext{a}{I.P.=\,ionization potential.} 
\tablenotetext{b}{Solar reference abundance for element M from Asplund et al.\,(2009).}
\tablenotetext{c}{Listed errors include uncertainties from the column-density
measurements of X and H\,{\sc i}.}
\tablenotetext{d}{Gas phase abundance for element M,
defined as [M/H]\,=\,log\,(M/H)\,$-$\,log\,(M/H)$_{\sun}$, derived from the
CLOUDY model. The listed errors include uncertainties from the column-density
measurements, the beam-smearing of the 21cm data, and the ionization correction
from the Cloudy modeling.} 
\tablenotetext{e}{Depletion value, defined as 
log $\delta$(M)=\,[M/H$]_{\rm MS}-$[S/H$]_{\rm MS}$.} 
\end{deluxetable*}

%

From the comparison between the predicted and measured 
ion column densities, together with the solar reference abundances
from Asplund et al.\,(2009), we obtain the following gas-phase abundances:
[N/H$]=-1.15\pm 0.06$,
[Si/H$]=-0.57\pm 0.06$,
[Fe/H$]=-0.86\pm 0.06$,
[S/H$]=-0.30\pm 0.04$,
[Al/H$]=-0.92\pm 0.08$,
[Ti/H$]=-1.62\pm 0.07$, 
[Ni/H$]\leq-0.81$, and
[Ca/H$]\leq0$.
The listed errors include the uncertainties from the 
column density measurements for the metal ions and 
H\,{\sc i} (Table 1), the systematic uncertainty due to the 
beam size of the 21cm measurements (Sect.\,3.3), and 
the systematic uncertainty for the ionization correction
from the Cloudy model (see above). The relative contributions
of these uncertainties to the total error budget are roughly
identical. Note that we here do not consider
the systematic errors that come with the solar reference
abundances (Asplund et al.\,2009; see Table 3, third row), because these
errors are irrelevant for the comparison between our results
and other abundance measurements, if the identical reference
values are used. All gas-phase abundances derived with Cloudy
are listed in Table 3, sixth row.

The measured S/H ratio of [S/H$]=-0.30\pm 0.04$ corresponds to an 
overall sulfur abundance in the Magellanic Stream toward 
Fairall\,9 of $0.50^{+0.05}_{-0.04}$ solar. The value for 
[S/H] is $0.25$ dex higher
than the value derived for the MS toward Fairall\,9
by Gibson et al.\,(2000) based on GHRS data. Several factors 
contribute to this discrepancy: (a) the somewhat higher ($+0.08$ dex)
S\,{\sc ii} column density that we derive from the COS data, 
(b) the slightly lower H\,{\sc i} column density ($-0.02$ dex) 
that we obtain from the GASS data, and (c) the substantially
higher ($0.15$ dex) solar reference abundance of sulfur 
(Asplund et al.\,2009) that we use to derive [S/H] compared to the
value from Anders \& Grevesse (1989) used by Gibson et al.\,(2000).

Remarkably, the sulfur abundance we derive here is 
{\it higher} than the present-day stellar sulfur abundances in the SMC
([S/H$]=-0.53\pm 0.15$) {\it and} in the LMC ([S/H$]=-0.42\pm 0.09$; 
Russell \& Dopita 1992), but matches the present-day interstellar 
S abundances found in H\,{\sc ii} regions in the Magellanic 
Clouds (Russell \& Dopita 1990). This interesting result will
be further discussed in Sect.\,5. Because of the large abundance 
scatter of sulfur in the Magellanic Clouds and 
because S appears to be underabundant 
compared to oxygen in the LMC and overabundant compared to oxygen 
in the SMC (Russel \& Dopita 1992), the measured (S/H) ratio alone 
does not provide a constraining parameter to pinpoint the origin of the 
gaseous material in the Stream toward Fairall\,9 in either SMC 
or LMC. 

From our Cloudy model it follows that all other elements listed above have 
gas-phase abundances (or abundance limits) that are lower than that of S. 
Next to the intrinsic nucleosynthetic MS abundance pattern, dust depletion is
another important effect that contributes to the deficiency of certain 
elements (e.g., Ti, Al, Fe, Si) in the gas phase. This aspect will be discussed in 
Sect.\,4.3. Note, again, that we do not attempt to constrain a single
value for log $U$ or log $n_{\rm H}$ with our Cloudy model, as our 
sightline passes a complex multiphase structure that is expected to span
a substantial {\it range} in gas densities and ionization parameters.

In this paper, we do not further analyze the gas phase that
is traced by the intermediate and high ions Si\,{\sc iii}, 
C\,{\sc iv} and Si\,{\sc iv}, which
probably is arising in the ionized boundary layer between the neutral
gas body of the Stream and the ambient hot coronal halo gas of the
Milky Way (see Fox et al.\,2010).
For completeness, we show in Fig.\,4 the velocity 
profiles of the available transitions of Si\,{\sc iii}, 
C\,{\sc iv} and Si\,{\sc iv}
in the COS spectrum of Fairall\,9. The 
velocity structure in these ions clearly is different from that
of the low ions, suggesting that they trace spatially 
distinct regions in the MS. The high-ion absorption 
toward Fairall\,9 will be included in a forthcoming
paper that will discuss those ions in several lines of
sight intersecting the Magellanic Stream 
(A.J. Fox et al.\,2014, in preparation).


\begin{figure}[t!]
\epsscale{1.0}
\plotone{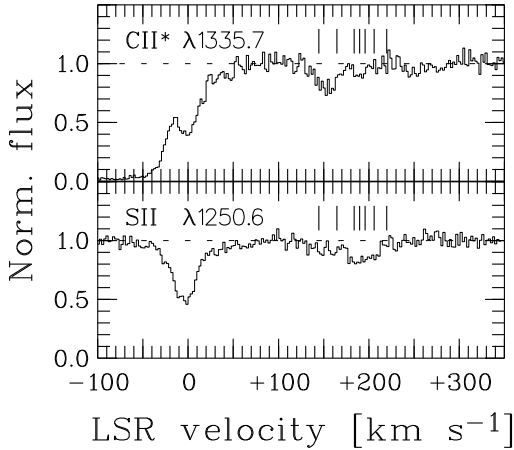}
\caption{
COS velocity profile of C\,{\sc ii}$^{\star}$ $\lambda 1335.7$ (upper panel).
C\,{\sc ii}$^{\star}$ absorption from the Magellanic Stream
is visible between $+120$ and $+200$ km\,s$^{-1}$. For comparison,
the velocity profile of S\,{\sc ii} $\lambda 1250.6$ is also
shown (lower panel). The component structure of the low ions is indicated
with the tic marks.}
\end{figure}

%

\subsection{Nitrogen abundance}

The measured gas phase abundance of nitrogen in the Magellanic Stream 
toward Fairall 9 is [N/H$]=-1.15\pm 0.06$, which is $0.85$ dex lower than 
the abundance of sulfur. In diffuse neutral gas the depletion of nitrogen into
dust grains is expected to be very small (Savage \& Sembach 1996). 
Consequently, the low nitrogen abundance in the Stream reflects the 
nucleosynthetic enrichment history of the gas and thus provides important 
information about the origin of the gas.

For the SMC, Russell \& Dopita (1992) derive a mean present-day 
stellar nitrogen abundance of [N/H$]=-1.20$. This value 
is close to the value derived for the MS toward 
Fairall\,9, although the gas in the Stream has not been enriched
with nitrogen since it was stripped off its parent galaxy.
The nitrogen abundances found in SMC H\,{\sc ii} regions and 
supernova remnants span a rather wide range of 
$-1.37\leq$[N/H$]\leq -0.95$ (Russell \& Dopita 1990). Thus, 
the observed nitrogen abundance in the MS toward Fairall\,9 
would be in accordance with an SMC origin of 
the gas only if the (mean) nitrogen abundance in the SMC 
has not increased substantially after the Stream was 
separated $\sim 1-2$ Gyr ago.

In the LMC, in contrast, the mean present-day stellar nitrogen 
abundance is found to be much higher ([N/H$]=-0.69$).
The N abundance range in LMC H\,{\sc ii} regions and supernova 
remnants is [N/H$]=-0.86$ to $-0.38$, the minimum [N/H] still being 
substantially higher than the nitrogen abundance in the 
Stream toward Fairall\,9. Therefore, an LMC origin of
the MS is plausible only if the LMC has substantially 
increased its nitrogen abundance during 
the last $\sim 2$ Gyr.

In Sect.\,5.1 we will further consider these interesting aspects 
and discuss the observed N abundance in the MS in the context of the 
enrichment history of the Magellanic Clouds.

\subsection{Dust depletion pattern}

Heavy elements such as Al, Si, Fe, Ca, Ti, Fe and Ni are known to be strongly
depleted into dust grains in interstellar gas in the Milky Way and 
other galaxies, while other elements such as S, O, N are 
only very mildly or not at all affected by dust depletion 
(e.g., Savage \& Sembach 1996; Welty et al.\,1997).

To study the depletion pattern of these elements in the 
Magellanic Stream we define the depletion values
from the relative abundance of each depleted element, M,
in relation to the abundance of sulfur. 
We define accordingly 
log $\delta$(M)=\,[M/H$]_{\rm MS}-$[S/H$]_{\rm MS}$.
Using this definition, the depletions values are
log $\delta$(Si$)=-0.27$,
log $\delta$(Fe$)=-0.56$,
log $\delta$(Ti$)=-1.32$,
log $\delta$(Al$)=-0.62$, and
log $\delta$(Ni$)\leq -0.51$ (see also Table 3, seventh row).
Such depletion values are typical for warm Galactic halo clouds,
as derived from UV absorption-line measurements (see Savage \& Sembach 1996;
their Fig.\,6).

Note that the depletion values (or limits) would be different, if
the {\it relative} chemical abundances of the depleted elements would 
differ from the relative chemical composition of the 
sun (Asplund et al.\,2009).
In view of the abundance patterns found in the LMC and SMC
(Russell \& Dopita 1990,1992), this is actually a likely scenario,
but no further conclusions can be drawn at this point
without knowing the exact intrinsic chemical composition of the 
MS gas toward Fairall\,9. 

Since the expected Ca\,{\sc ii} column density in MS gas is expected
to depend strongly on the local gas density and ionization parameter
(Fig.\,6), the depletion value of Ca cannot be tightly constrained,
but may lie anywhere in the range log $\delta$(Ca$) \geq -1.58$.
Under typical interstellar conditions, Ca is strongly depleted
into dust grains with large depletion values that are similar to those
of other elements that have condensation temperatures above
$T=1500$ K (e.g., Ti; Savage \& Sembach 1996). 
If one {\it assumes} that
log $\delta$(Ca$) \approx$\,log $\delta$(Ti) for the MS 
toward Fairall\,9, then Fig.\,6 would imply that 
the average density in the neutral gas in the Stream 
in this direction is log $n_{\rm H}\approx 0$ or 
$n_{\rm H}\approx 1$ cm$^{-3}$.
This density, together with the neutral gas column density,
would imply a characteristic thickness of the absorbing 
neutral gas layer of $d=N($H\,{\sc i}$)/n_{\rm H}\approx 30$ pc. 

At first glance, one may regard this as a relatively small value. 
The complex velocity-component structure of the 
absorber indicates, however, that the absorber is composed
of several small cloudlets along the line of sight that
(together) are extending over a spatial range that is 
larger than the value indicated by the {\it mean} thickness. 
In view of the relatively high metal abundance derived 
for the kinematically complex absorber toward Fairall\,9,
the presence of a population of metal-enriched gas 
clumps at small scales in the MS is not an unlikely scenario.

\subsection{Physical conditions in the H$_2$ absorbing region}

As discussed in Sect.\,3.1, the UVES spectrum of Fairall\,9 shows 
the presence of Na\,{\sc i}, Ca\,{\sc ii}, and possibly Ti\,{\sc ii}
in component 4. In the same component, the detected H$_2$ absorption 
is expected to arise (Sect.\,3.3), indicating the presence 
of a cold, dense (and predominantly neutral) gas phase in the MS 
in this direction. In the following, we want
to combine the information from the different instruments to 
explore the physical and chemical conditions in the H$_2$ absorbing
gas phase in the MS toward Fairall\,9.

\subsubsection{Gas density}

While our Cloudy model does not provide any relevant information 
on the {\it local} gas densities in the individual subcomponents, 
we can use the observed molecular hydrogen fraction in the gas to 
constrain $n_{\rm H}$ in the H$_2$ absorbing component 4, assuming 
that the abundance of H$_2$ relative to H\,{\sc i} is governed by 
a formation-dissociation equilibrium. In a formation-dissociation 
equilibrium the neutral to 
molecular hydrogen column density ratio in an interstellar  
gas cloud is given by

\begin{equation}
\frac{N({\rm H\,I})}{N({\rm H}_2)} =
\frac{\langle k \rangle \,\beta}{R\,n({\rm H\,I})\,\phi},
\end{equation}

\noindent
where $\langle k \rangle \approx  0.11$ is the
probability that the molecule is dissociated after photo absorption,
$\beta$ is the photo-absorption rate per second within the
cloud, and $R$ is the H$_2$ formation rate on dust grains
in units cm$^{3}$\,s$^{-1}$. 
For low molecular hydrogen fractions we can 
write $n_{\rm H} = n({\rm H\,I})+2n({\rm H}_2) 
\approx n({\rm H\,I})$. The parameter $\phi \leq 1$ 
in equation (1)
describes the column-density fraction of the H\,{\sc i} 
that is physically related to the H$_2$ absorbing gas,
i.e., the fraction of the neutral gas atoms that can
be transformed into H$_2$ molecules (see Richter et al.\,2003
for details). 
Our absorption modeling of the undepleted low ions 
S\,{\sc ii} and N\,{\sc i} indicates that component
4 contains $\sim 15-30$ percent of neutral gas column,
so that we assume $\phi=0.15-0.30$ in lack of any
more precise information. 

For interstellar clouds that are optically thick in
H$_2$ (i.e., for log $N$(H$_2)\gg 14$) H$_2$ line self-shielding
needs to be considered in equation (1), because the self-shielding 
reduces the photo-absorption rate ($\beta$) in the cloud interior.
Draine \& Bertoldi (1996) find that the H$_2$ self-shielding 
can be expressed by the relation $\beta=S\,\beta_0$,
where $S=(N_{\rm H_2}/10^{14}$cm$^{-2})^{-0.75}<1$ is the
self-shielding factor and $\beta_0$ is the photo absorption rate 
at the edge of the cloud. The parameter $\beta_0$ is directly 
related to the intensity of the ambient UV radiation field
at the edge of the H$_2$-bearing clump. Compared to the 
Milky Way disk, where UV bright stars dominate 
the mean photo-absorption rate of
$\beta_{\rm 0,MW}=5.0\times 10^{-10}$ s$^{-1}$ 
(e.g., Spitzer 1978), the value for $\beta_0$
in the Magellanic Stream is expected to be substantially 
smaller. This is because the solid angle of the Milky
Way disk is relatively small at 50 kpc distance and
the contribution of the extragalactic UV background to
$\beta_{\rm 0,MS}$ is expected to be (relatively) small, too.
From the model by Fox et al.\,(2005) follows that the UV
flux between 900 and 1100 \AA\, is reduced by a factor
of $160$ compared to the mean flux within the Milky Way disk,
so that we assume $\beta_{\rm 0,MS}=3.1\times 10^{-12}$ s$^{-1}$.
With a total H$_2$ column density of $8.5\times 10^{17}$ cm$^{-2}$
in the Stream the self-shielding factor $S$ becomes $1.13 \times
10^{-3}$, so that the photo-absorption rate in the cloud core is
estimated as $\beta = 3.5\times 10^{-15}$ s$^{-1}$.
For the H$_2$ grain formation rate in the Magellanic 
Stream we adopt the value derived for the SMC 
based on \emph{FUSE} H$_2$ absorption-line data 
(Tumlinson et al.\,2002), i.e.,
$R_{\rm MS} = 3 \times 10^{-18}$ cm$^{3}$\,s$^{-1}$. This 
value is 10 times smaller than $R$ in the
disk of the Milky Way (Spitzer 1978).

If we solve equation (3) for $\phi n_{\rm H}\,=\phi n({\rm H\,I})$
and include the values given above, we obtain
a density of $\phi n_{\rm H}\approx 1$ cm$^{-3}$ or 
$n_{\rm H}\approx 4-8$ cm$^{-3}$ for $\phi=0.15-0.30$. These
densities are very close to the density derived 
for the H$_2$ absorbing gas in the Leading Arm of the 
MS toward NGC\,3783
($n_{\rm H}\approx 10$ cm$^{-3}$; Sembach et al.\,2001). 
Note that H$_2$ absorption has also been detected 
in the Magellanic Bridge (Lehner 2002).

\subsubsection{Gas temperature}

%

\begin{figure}[t!]
\epsscale{1.2}
\plotone{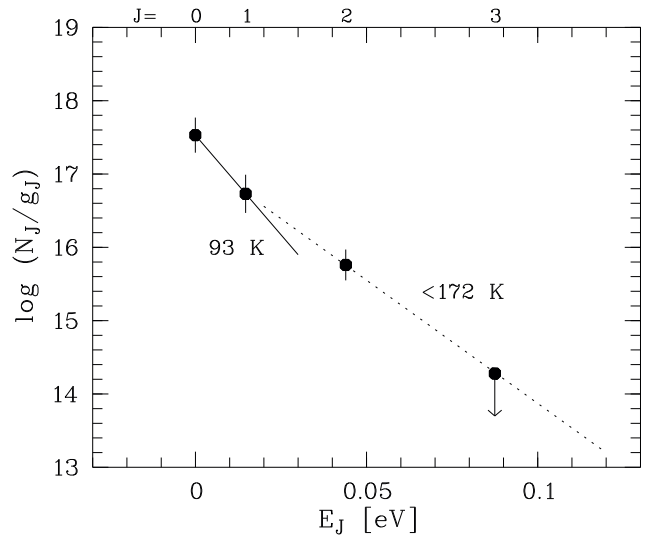}
\caption{Rotational excitation of H$_2$ in the Magellanic
Stream. The logarithmic H$_2$ column density
for each rotational state, log $N(J)$, divided by its statistical weight,
$g_J$, is plotted against the rotational excitation energy, $E_J$.
The fits for the excitation temperatures
(assuming a Boltzmann distribution) for $J=0,1$ and
$J=2,3$ are indicated with the solid and dashed line,
respectively.
}
\end{figure}

%

In Fig.\,8 we plot the measured logarithmic H$_2$ column density 
for each rotational state, log $N(J)$, divided by its statistical weight,
$g_J$, against the rotational excitation energy, $E_J$.
Rotational excitation energies and statistical weights have been
adopted from the compilation of Morton \& Dinerstein (1976).
The data points in Fig.\,8 can be fit by a Boltzmann
distribution (i.e, a straight line in this plot),
where the slope characterizes the (equivalent)
excitation temperature for the rotational levels considered
for the fit. For the two rotational ground states $J=0,1$
the Boltzmann distribution thus is given by the equation
$N(1)/N(0)=g_1/g_0\,{\rm exp}\,(-E_{01}/kT_{01})$.
Using this equation, together with our measured column
densities for $J=0,1$, we find an excitation temperature
of $T_{\rm 01}=93^{+149}_{-39}$ K (Fig.\,8 solid line).
Because the two rotational ground states most likely are
collisionally excited, $T_{\rm 01}$ represents a robust
measure for the kinetic temperature in the H$_2$ absorbing gas
(e.g., Spitzer 1978); however, because of the saturation of the
H$_2$ lines and the resulting relatively large errors for
$N(0,1)$, the uncertainty for $T_{\rm 01}$ is 
substantial.
In a similar manner, we derive for the rotational states
$J=2,3$ an upper limit for the 
equivalent Boltzmann temperature of
$T_{\rm 23}\leq 172$K (Fig.\,8 dashed line). This higher
excitation temperature for $J\geq2$ reflects excitation
processes other than collisions, such as UV pumping
and H$_2$ formation pumping (Spitzer 1978). However, 
the relatively low temperature limit of $T_{\rm 23}\leq 172$ K 
reflects the low intensity of the ambient 
dissociating UV radiation field at the position of the MS.

The above given excitation temperatures deviate from
our earlier estimates  (Richter et al.\,2001a). This is not
surprising, however, since the values for $N(J)$ have 
substantially changed, too (see Sect.\,3.3).
The derived value of $T_{\rm 01}=93$ K lies in a 
temperature regime
that is typical for diffuse H$_2$ gas in the disk of the
Milky Way and in the Magellanic Clouds (e.g., Savage et al.\,1977;  
Tumlinson et al.\,2002; de Boer et al.\,1998; Richter et al.\,1998;
Richter 2000).

From the density and temperature estimate in the H$_2$ absorbing
gas we now are able to provide a direct estimate of the thermal 
pressure, $P/k=nT$, in the cold neutral gas in the MS. With 
$n_{\rm H}= 8$ cm$^{-3}$ and $T=93$ K we obtain
a pressure of $P/k=744$ cm$^{-3}$\,K. This value is 
in good agreement with previous estimates for the 
thermal pressure in the cold neutral phase of the 
MS from H\,{\sc i} 21cm
measurements (e.g., Wakker et al.\,2002). 

The thickness of the H$_2$ absorbing structure in the MS
is $d_{\rm H_2}=\phi\,N$(H\,{\sc i}$)/n_{\rm H}
\approx 0.6-1.2$ pc. This
small dimension of the absorbing gas clump
explains the very small velocity dispersion 
($b=1.8$ km\,s$^{-1}$; see Table 1) that is measured for
this component.

\subsection{Electron density and C$^{+}$cooling rate 
from C\,{\sc ii}$^{\star}$ absorption}

In Fig.\,7 (upper panel) we show the LSR velocity profile of 
the C\,{\sc ii}$^{\star}$
$\lambda 1335.7$ line, based on the COS data. Weak absorption
is visible at MS velocities, but the component structure is 
different from that of the other low ions (e.g., S\,{\sc ii}; 
Fig.\,7, lower panel).
Because we could not identify any other (e.g., intergalactic) 
origin for this absorption feature, we assume that it is 
caused by C\,{\sc ii}$^{\star}$ in the Magellanic Stream.
The C\,{\sc ii}$^{\star}$ feature is characterized by a stronger and broader
absorption peak near $+150$ km\,s$^{-1}$, falling
together with components 1 and 2 defined in Table\,1, and 
another weak and narrow component that is seen near $+190$ km\,s$^{-1}$,
most likely associated with component 4, in which H$_2$, Na\,{\sc i},
and Ca\,{\sc ii} is detected. Therefore, the relative strengths
of the C\,{\sc ii}$^{\star}$ components appear to be inverted from 
those of the ground-state species. This indicates multi-phase gas,
in which the ionization fractions and electron densities vary among the
different absorption components.

Under typical interstellar conditions, the relative population of the 
fine-structure levels of ionized carbon (C$^{+}$) are governed by the balance
between collisions with electrons and the radiative decay
of the upper level into the ground state $2s^2 2p ^2P_{1/2}$. 
The C\,{\sc ii}$^{\star}$ $\lambda 1335.7$ transition arises from 
the $2s^2 2p ^2P_{3/2}$ state, which has an energy of $\sim 8\times 10^{-5}$ 
eV above the ground state. Measurements of the column-density 
ratios $N$(C\,{\sc ii}$^{\star}$)/$N$(C\,{\sc ii}) thus can be 
used to estimate the electron densities $n_{\rm e}$ in different 
interstellar environments, including Galactic high-velocity clouds 
(HVCs) and circumgalactic gas 
structures (e.g., Zech et al.\,2008; Jenkins et al.\,2005).

For the C\,{\sc ii}$^{\star}$ absorption in the MS toward 
Fairall\,9 we measure a total column density of 
log $N$(C\,{\sc ii}$^{\star})=13.35\pm 0.07$ for the velocity
range $v_{\rm LSR}=120-220$ km\,s$^{-1}$ using the AOD
method. For the weak absorption associated with component 4
we derive log $N$(C\,{\sc ii}$^{\star})=12.63\pm 0.03$ (AOD). 
The C\,{\sc ii} $\lambda 1334.5$ absorption in the MS is fully saturated 
and thus does not provide a constraining limit on $N$(C\,{\sc ii})
(see Table 1). We therefore use S\,{\sc ii} as a proxy, because
the S\,{\sc ii}/C\,{\sc ii} ratio is expected to be constant 
over a large density range (Fig.\,6) and both elements
are not expected to be depleted into dust grains in the MS. Using 
a solar (S/C) ratio (Asplund et al.\,2009), we estimate from the values listed
in Table 1 that the total C\,{\sc ii} column density in the MS 
is log $N_{\rm tot}$(C\,{\sc ii}$)\approx 16.10$ and 
the column density in component 4 is 
log $N_{4}$(C\,{\sc ii}$)\approx 15.30$.

Because the C$^{+}$ fine-structure population depends 
strongly on the gas temperature (see, e.g., Spitzer 1978),
we do not attempt to estimate a mean value for $n_{\rm e}$ from 
$N_{\rm tot}$(C\,{\sc ii}$^{\star})$/$N_{\rm tot}$(C\,{\sc ii}).
We know that the temperature in this multi-phase
absorber spans a large range between $\sim 100$ K and
probably few $1000$ K, and therefore an
estimate for $\langle n_{\rm e} \rangle$ would be meaningless.
Instead, we concentrate on component 4, for which we 
know the gas temperature and density from the analysis 
of the H$_2$ rotational excitation (Sect.\,4.4). 
Keenan et al.\,(1986) have calculated electron excitation
rates for the C$^{+}$ fine-structure transitions for a 
broad range of physical conditions in interstellar gas.
Using our temperature and density estimates for the 
gas in component 4 ($T=93$ K and $n_{\rm H}=4-8$ 
cm$^{-3}$) together with their predicted population
rate ratios ($2s^2 2p ^2P_{3/2}$/$2s^2 2p ^2P_{1/2}$)
for $T=100$ K and $n_{\rm H}=5-10$ cm$^{-3}$ (their Fig.\,2), 
the measured column-density ratio 
log[$N$(C\,{\sc ii}$^{\star}$)/$N$(C\,{\sc ii})$]=-2.67$
implies that the electron density in component 4
is small compared to $n_{\rm H}$, namely
$n_{\rm e}\leq 0.05$ cm$^{-3}$. This low value
for $n_{\rm e}$ is in excellent agreement with
expectations for a stable cold, neutral medium
in gas with subsolar metallicities (Wolfire et al.\,1995).

Having a robust measure for $N_{\rm tot}$(C\,{\sc ii}$^{\star}$), 
it is also possible to estimate the C$^{+}$ cooling rate in
the Magellanic Stream toward Fairall\,9. 
Because ionized carbon is a major cooling agent of interstellar 
gas in a wide range of environments (e.g., Dalgarno \& McCray 1972),
the C$^{+}$ cooling rate is an important quantity that governs the 
thermal state of diffuse gas inside and outside of galaxies. 
For the hydrogen and electron densities in the Magellanic Stream 
(see above) the C$^{+}$ cooling rate is governed predominantly 
by spontaneous de-excitations, while collisional de-excitations 
can be neglected (e.g., Spitzer 1978). 
Following Lehner, Wakker \& Savage (2004), 
the C$^{+}$ cooling rate per neutral hydrogen atom can be estimated as
$l_C=2.89 \times 10^{-20}\,N($C$^{+})/N($H\,{\sc i}$)$\,erg\,s$^{-1}$.
If we adopt log $N$(H\,{\sc i}$)=19.95$ and log $N$(C\,{\sc ii}$^{\star})=13.35$
we derive a mean (sightline average) C$^{+}$ cooling rate per neutral hydrogen atom
of log $l_C=-26.14$ for the MS toward Fairall\,9. This value is almost one
order of magnitude lower
than the cooling rate derived for the one-tenth-solar metallicity HVC Complex C
(log $l_C\approx -27$), but is relatively close to the values derived for solar-metallicity 
clouds in the Milky Way disk and in the lower Galactic halo
(log $l_C\approx -26$; see Lehner, Wakker \& Savage 2004, their Table\,4).
This result underlines the importance of metal cooling for the thermal state
of metal-rich circumgalactic gas absorbers in the local Universe.

%

\section{Discussion}

\subsection{Enrichment history of the Magellanic Stream}

\subsubsection{Alpha and nitrogen abundances}

One major result of our study is the surprisingly high
metallicity of the MS toward Fairall\,9. From the 
measured sulfur abundance of [S/H$]=-0.30\pm 0.04$ 
follows that the $\alpha$ abundance in the Stream in
this direction ($l=295, b=-58$) is as high as $0.5$ solar, which is 
$\sim 5$ times higher than the $\alpha$ abundance derived
for other sightlines passing the MS toward 
NGC\,7469 ($l=347, b=-64$), RBS\,144 ($l=299, b=-66$), and 
NGC\,7714 ($l=88, b=-56$; Fox et al.\,2010, 2013).
Our detailed analysis of the ionization conditions in the 
gas, the comparison between H\,{\sc i} 21cm measurements
from different radio telescopes with different beam sizes,
and the analysis of the H\,{\sc i} Ly\,$\alpha$ absorption
toward Fairall\,9 (see Appendix) do not provide any
evidence, that this high sulfur abundance could be a result 
from the various systematic errors that come along with
our analysis (see also the discussion in Gibson et al.\,2000 
on this topic).
We thus are forced to conclude that the measured high sulfur 
abundance in the gas reflects the true chemical composition
of the Stream in this direction.

As a guide to the following discussion, we have plotted 
in Fig.\,9 the derived MS abundances
(black filled circles) together with
the SMC and LMC present-day stellar and nebular 
abundances (Russel \& Dopita 1992; Hughes et al.\,1998; 
red (SMC) and green (LMC) filled circles).
With a sulfur abundance of $0.5$ solar, the MS toward Fairall\,9
exhibits an $\alpha$ abundance that is {\it higher} than the average
present-day $\alpha$ abundance in both SMC and LMC 
($\sim 0.3$ and $\sim 0.4$ solar, respectively; Russell \& Dopita 1992).
This finding has profound implications for our understanding of the
origin of the Magellanic Stream and its enrichment history.

Early tidal models suggest that the MS was stripped from the 
Magellanic Clouds $\sim 2$ Gyr ago (Gardiner \& Noguchi 1996). More
recent models, that take into account that the Magellanic Clouds 
possibly are on their first pass through the Milky Way halo
(see, e.g., Kallivayalil et al.\,2013),
have included other relevant mechanisms that would 
help to unbind the gas, such as blowouts and tidal resonances
(Connors et al.\,2006; Nidever et al.\,2008; Besla et al.\,2010).
One would expect that 
the Stream's chemical composition must reflect that of its 
parent galaxy at the time, when the gas was separated from the 
Magellanic Clouds, since the MS does not contain stars that
would increase its metallicity since then. Following the 
well-defined age-metallicity relations of SMC and LMC (Pagel \&
Tautvaisiene 1998; Harris \& Zaritsky 2004; 2009), 
one would expect that the Stream's present-day
$\alpha$ abundance is at most $\sim 0.2$ solar if originating
in the SMC and $\sim 0.25$ solar if originating
in the LMC. While the observed MS abundances toward RBS\,144,
NGC\,7469, and NGC\,7714 are in line with these abundance limits
(and actually favour an SMC origin for this part of the
Stream; see Paper\,I), the sulfur abundance in the
MS toward Fairall\,9 is at least twice as high as expected
from a simple tidal model that assumes a homogeneous pre-enrichment 
of the SMC/LMC gas (e.g., Dufour 1975) before the Stream was stripped off.

A second important finding of our study, that adds to this puzzle,
is the relatively low nitrogen abundance in the MS toward
Fairall\,9. The [N/S] ratio ($=[$N/$\alpha$] ratio) is $-0.85$ dex,
which is very low for the relatively high $\alpha$ abundance of 
[$\alpha$/H]$=-0.3$ when compared to nearby extragalactic H\,{\sc ii}
regions and high-redshift damped Lyman $\alpha$ (DLA) systems
(see Pettini et al.\,2008; Jenkins et al.\,2005, and references
therein). The $\alpha$-process elements O, S, and Si are believed
to be produced by Type II supernovae from massive progenitor stars,
while the production of N, as part of the CNO cycle in stars 
of different masses, is less simple (and not yet fully understood). 
The so-called ``primary'' nitrogen production occurs when the
seed elements C and O are produced within the star during the
helium burning phase, while for ``secondary'' nitrogen 
production these seed elements were already present when the star
condensed out of the ISM (Pettini et al.\,2002;
Henry \& Prochaska 2007). Primary N is believed to be produced
predominantly by stars of intermediate masses on the asymtotic
giant branch (e.g., Henry et al.\,2000). Because of the longer 
lifetime of intermediate-mass stars, primary N therefore is expected 
to be released into the surrounding ISM with a time delay of 
$\sim 250$ Myr compared to $\alpha$ elements. At low metallicities
less than $\sim 0.5$ solar (i.e., at LMC/SMC metallicities 
$\sim 1-2$ Gyr ago, when the MS was separated from its parent galaxy)
the nitrogen production is predominantly primary (see Pettini
et al.\,2008).

The low [N/S] and high [S/H] ratios observed in the MS toward 
Fairall\,9 therefore indicate an abundance pattern that is dominated by 
the $\alpha$ enrichment from massive stars and Type II supernovae,
while only very little (primary) nitrogen was deposited into
the gas. 

%

\begin{figure}[t!]
\epsscale{1.2}
\plotone{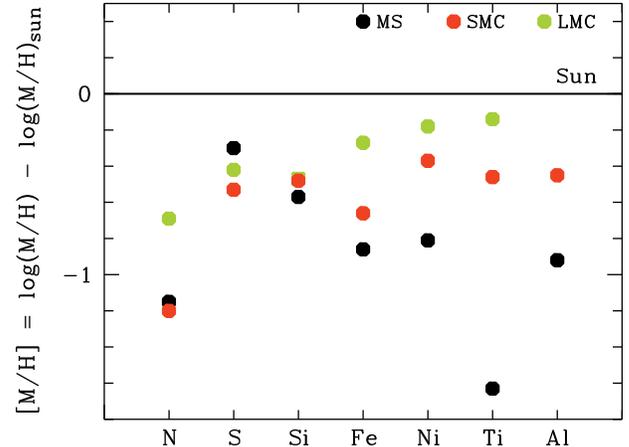}
\caption{Comparison of present-day metal abundances in the
SMC (red dots), LMC (green dots), and the Magellanic Stream
toward Fairall\,9 (black dots). For SMC and LMC the metallicities
are derived from stellar and nebular abundances (from Russell \&
Dopita 1992; Hughes et al.\,1998).
}
\end{figure}

%

\subsubsection{Origin of the gas}

The most plausible scenario that would explain 
such an enrichment history, is, that the gas that later became 
the Stream was {\it locally} enriched in the Magellanic Clouds 
$\sim 1-2$ Gyr ago with $\alpha$ elements by several supernova explosions in a star 
cluster or OB association, and then separated/stripped from the stellar body 
of the parent galaxy {\it before} the primary nitrogen was 
dumped into the gas and the metals could mix into the ambient
interstellar gas. Eventually, the supernova explosions 
may have pushed away the enriched material from the stellar disk
so that the gas was already less gravitationally bound at the
time its was stripped and incorporated into the Magellanic Stream.
In this scenario, it is the present-day N abundance in the MS that 
defines the metallicity floor of the parent galaxy and thus
provides clues to the origin of the gas.

Chemical evolution models of the Magellanic Clouds suggest that the 
mean metallicities of SMC and LMC $1-2$ Gyr ago were 
$0.2-0.3$ dex lower than today (Pagel \& Tautvaisiene 1998;
Harris \& Zaritsky 2004, 2009). 
Therefore, and because of the substantially higher mean N abundance
in the LMC compared to the SMC (see Fig.\,9), the low 
N abundance measured in the Stream toward Fairall\,9 
favours a SMC origin of the gas. An LMC origin is also possible,
however, because the scatter in the 
present-day N abundances in the LMC is large (see Sect.\,4.2)
and the age-metallicity relations from Pagel \& Tautvaisiene (1998)
and Harris \& Zaritsky (2004,2009) may not apply to nitrogen
because of its special production mechanism. In fact, considerung 
the position of the Fairall\,9 sightline (see Paper\,I, Fig.\,1) and 
the radial velocity of the MS absorption in this direction,
the bifurcation model of Nidever et al.\,(2008) {\it predicts} that the
high-velocity gaseous material toward Fairall\,9 is part
of the LMC filament of the Stream und thus should have a different
chemical composition than the SMC filament traced by the other 
MS sightlines (Paper\,I). Another aspect that may be of relevance
in this context is the fact that the Fairall\,9 sightline 
lies only 14.3\degr\ on the sky from the SMC. The relative high
metal-abundance in the Stream toward Fairall\,9 thus may reflect
the increasing importance of continuous ram-pressure stripping of 
metal-enriched SMC gas as the MCs get closer to the Milky Way.

While the above outlined enrichment scenario appears feasible to 
explain qualitatively the observed trend seen in the abundance pattern 
in the MS toward Fairall\,9, the question arises, how many massive 
stars and subsequent supernova explosions would then be required
to lift the [S/N] and [S/H] ratios to such a high level. To 
answer this question, one would first need to have an idea about
volume and mass of the gas that is enriched in this manner.
The fact that Fox et al.\,(2010, 2013) have determined a
much lower $\alpha$ abundance in the Stream of $\sim 0.1$ solar
toward three other MS sightlines using similar data 
possibly suggests, that the high metallicity and low [N/S] ratio
toward Fairall\,9 represents a relatively local 
phenomenon.

In Sect.\,4.3, we have estimated a thickness of the absorbing 
neutral gas layer in the Stream of $d\approx30$ pc from 
the mean gas density ($n_{\rm H}=1.0$ cm$^{-3}$) and 
the total column density, assuming that the dust depletion values 
of Ti and Ca are similiar in the gas. In the following, we 
regard this value as a realistic {\it lower} limit for the true 
physical size of the absorbing gas region. If we adopt the 
SN yields from Kobayashi et al.\,(2006) and assume a spherical 
symmetry for the absorbing gas region with a diameter and density as
given above, we calculate that only a handful of massive 
stars would be required to boost the $\alpha$ abundance 
from initially $0.1$ solar to $0.5$ solar 
in such a small volume of gas. A much higher 
number of high-mass stars would be required, however,
if the volume to be enriched would be substantially larger, as we
outline in the following.
Because the total mass of a spherical gas cloud can be written 
as the product of cloud volume and gas density 
($M\propto Vn_{\rm H}\propto d^3n_{\rm H}$), and because we use the 
relation $d=N_{\rm H}/n_{\rm H}$ to estimate $d$, we can write 
$M\propto n_{\rm H}^{-2}$ for a fixed (measured) hydrogen column 
density in the gas. Therefore, if the mean gas density in the 
MS toward Fairall\,9 would be $0.1$ cm$^{-3}$ rather than the
assumed $1.0$ cm$^{-3}$, then the diameter of the cloud 
would be 300 pc instead of
30 pc, but it would require 100 times more massive stars
to enrich this large volume to the level required. 
In view of the stellar content of the most massive star
forming regions in the Magellanic Clouds (e.g., 30\,Doradus;
Melnick 1985) and the possible presence of a major star burst 
in the MCs at the time when the Magellanic Stream was formed 
(Harris \& Zaritsky 2009; Weisz et al.\,2013), a size 
of a few dozen up to a few hundred pc together with the 
proposed enrichment scenario is fully consistent with 
our understanding of the star-formation history of the 
Magellanic Clouds.

In summary, the observed abundance pattern in the MS toward
Fairall\,9 suggests the presence of a high-metallicity gas filament
in the Stream in this direction, possibly originating in a
region with enhanced star-formation activity in the Magellanic Clouds
$\sim 1-2$ Gyr ago. Our study indicates that the chemical composition
and thus the enrichment history of the MS seems to be more complex
than previously thought. The Stream possibly is composed
of different gas layers that have different chemical
compositions and that originate from different regions
with local star-formation histories within the Magellanic Clouds.
Our findings therefore support (but do not require) the idea that 
there is a metal-enriched filament in the Stream toward Fairall\,9 that
originates in the LMC (but not the SMC), as concluded by 
Nidever et al.\,(2008) from a systematic Gaussian decomposition
of the H\,{\sc i} velocity profiles of the LAB 21cm all-sky
survey (Kalberla et al.\,2005). To further explore this
scenario it will be of great importance to identify other
sightlines that pass the supposed LMC filament in the Stream 
with background QSOs that are bright enough to be observed with
\emph{HST}/COS. Such observations would also 
be helpful to investigate the $\alpha$ abundances 
in the LMC filament based on oxygen rather than sulfur and
to further explore element ratios that could help 
to constrain the enrichment history of the gas.

Interestingly, the MS absorber toward Fairall\,9 is not the 
only example of a circumgalactic gas structure in the nearby 
Universe that exhibits a high overall metallicity together
with a low N/$\alpha$ abundance ratio. For example, 
Jenkins et al.\,(2005) have measured [$\alpha$/H]$=-0.19$
and [N/$\alpha]<0.59$ in an intervening Lyman-limit system (LLS) 
at $z=0.081$ toward the quasar PHL\,1811. This absorption system
appears to be associated with two nearby spiral galaxies at
impact parameters $\rho=34\,h_{70}^{-1}$ kpc and 
$\rho=87\,h_{70}^{-1}$ kpc.
It possibly represents gaseous material that has been ejected
or stripped from these galaxies (or other galaxies nearby) and
thus possibly has an origin that is very similar to that of the 
Magellanic Stream. Note that low [N/$\alpha$] ratios are
also found in low-metallicity HVCs, such as Complex C (e.g., 
Richter et al.\,2001b). Clearly, a systematic study of N/$\alpha$ 
ratios in circumgalactic metal absorbers at low $z$ with 
\emph{HST}/COS would be an important project to investigate 
whether similar abundance patterns are typical for circumgalactic
gaseous structures in the local Universe. 

\subsection{Physical conditions and small-scale structure in the gas}

In addition to the very interesting chemical properties, the combined 
UVES/COS/\emph{FUSE} /GASS/ATCA data set provides a deep insight into the physical 
conditions in the Stream toward Fairall\,9. The data show a complex 
multiphase gas structure that possibly spans a large range in gas 
densities and ionization conditions.

The presence of H$_2$ absorption toward Fairall\,9 
and NGC\,3783 (Sembach et al.\,2001) indicate that the MS and its 
Leading Arm hosts a widespread (because of the 
large absorption cross section/detection rate) cold neutral
gas phase, possibly structured in a large amount of small,
dense clumps or filaments. Assuming a total area of the 
MS and its Leading Arm of $\sim 1500$ deg$^2$, and
a typical size for these clumps of a few pc, there 
could be millions of these dense structures in the 
neutral gas body of the Stream, if the two H$_2$ detections
toward Fairall\,9 and NGC\,3783 reflects the true 
absorption-cross section of this gas phase throughout the
Stream's neutral gas body.

The fact that the MS can maintain significant amounts of H$_2$ 
at moderate gas densities ($n_{\rm H}=1-10$ cm$^{-3}$)
probably is a result of the relatively low intensity of 
the dissociating UV radiation at the Stream's location due
to the absence of local UV sources (see also Fox et al.\,2005).
Note that the low excitation temperature of H$_2$ for $J\geq 2$ 
(see also Sembach et al.\,2001) provides independent evidence for a
relatively weak ambient UV radiation field, and thus supports
this scenario. In the MS, the low dissociation rate is expected to
overcompensate the low formation rate (because of the lower 
metallicity and dust content compared to the Milky Way ISM),
thus resulting in molecular gas fractions that are relatively 
high for the given total gas column. With a grain-formation 
rate of $R_{\rm MS} = 3 \times 10^{-18}$ cm$^{3}$\,s$^{-1}$ and
a density of $n_{\rm H}=5$ cm$^{-3}$ (see Sect.\,4.4), the 
H$_2$ formation time is long, $t_{\rm form}=
(R_{\rm MS}n_{\rm H})^{-1}\approx 2$ Gyr, which matches the
tidal age of the Stream. Therefore, it seems most likely 
that the H$_2$ (or at least some fraction of it) has already
formed in the parent galaxy and then survived the subsequent 
tidal stripping.
A similar conclusion was drawn from the H$_2$ observations in the Leading
Arm of the MS toward NGC\,3783 (Sembach et al.\,2001).
Elaborating our idea, that the gas was locally enriched by massive stars 
shortly before the Stream was separated from its parent galaxy, one 
could imagine the presence of relatively dense, compressed shells and 
fragments that formed the seed structures for the formation 
of H$_2$ that then were carried along with the Stream's gaseous
body.

Independent constraints on the physical conditions in the gas
come from the observed Na\,{\sc i}/Ca\,{\sc ii} ratio.
We measure a Na\,{\sc i}/Ca\,{\sc ii} 
column-density ratio of $0.23$ in component 4, where the
H$_2$ is expected to reside. In the Milky Way ISM, such a ratio
is typical for a dust-bearing warm neutral medium (WNM), where 
$n_{\rm H}\leq 10$ cm$^{-3}$ and $T=10^2-10^4$ K, typcially,
and Ca\,{\sc ii} serves a trace species (e.g., Crawford 1992).
In such gas, and {\it without} dust depletion of Ca and Na, a nearly 
constant Na\,{\sc i}/Ca\,{\sc ii} of $0.025$ ratio would be 
expected from detailed ionization models of these ions 
(see Crawford 1992; Welty et al.\,1996; Richter et al.\,2011).
If these numbers would apply also to the conditions in the MS, 
they would indicate relative dust depletions of Ca and Na
of log $\delta$(Ca$)-$\,log $\delta$(Na$)\approx 0.9$, or
log $\delta$(Ca$)\approx0.9$ for log $\delta$(Na$)=0$. These
values are in very good agreement with the dust-depletion 
estimates from our Cloudy model discussed in Sect.\,4.3.

\subsection{Relevance to intervening QSO absorbers}

The Magellanic Stream represents a prime example for 
a high-column density circumgalactic tidal gas stream
in the local Universe. If the Stream would be seen as 
QSO absorber, it would be classified as LLS, sub-DLA 
(sub-Damped Lyman $\alpha$ absorber), or DLA, 
depending on the position of the sightline passing through
the gas. The H\,{\sc i} column-density maps presented in 
Fig.\,1 provides an estimate on the 
covering fractions of these column-density regimes
in the MS in the direction of Fairall\,9.
The results from our multi-sightline campaign to study
the properties of the MS therefore are also of relevance 
for the interpretation of interevening absorption-line 
systems at low redshift.

\subsubsection{Inhomogeneity of absorbers} 

The first important result from our 
study that is of relevance for QSO absorption-line studies is, 
that the physical conditions {\it and} the
chemical composition appear to vary {\it substantially} 
within the Stream.

If such inhomogeneities were typical for tidal gas 
filaments around galaxies, then the interpretation of absorption 
spectra from circumgalactic gas structures around more distant 
galaxies (in terms of metallicity, molecular content, 
physical conditions, gas mass, origin, etc.) would be 
afflicted with large systematic uncertainties. In fact,
most studies that aim at exploring the connection between
galaxies and their surrouding circumgalactic gas are
limited to single sightlines that pass the galaxy 
environment at a random impact parameter (e.g., 
Thom et al.\,2011; Tumlinson et al.\,2011; Ribaudo et al.\,2011).
Physical and chemical properties derived from single-sightline
studies, however, may not be representative at all for the
conditions in the general gaseous environment
(in the same way as conditions in the MS along the Fairall\,9 
sightline are {\it not} representative for the Stream as
a whole). 
The inhomogeneous chemical composition of the Stream implies
that the metals possibly are not well mixed in the gas.
Former abundance variations in the parent galaxy
due to local star-formation events thus may have been frozen
into the Stream's spatial metal distribution. This would be not
too surprising, however, since the main processes that stir up and mix 
the interstellar gas within galaxies (i.e., supernova explosions,
stellar winds) cannot take place in the Stream simply due to the lack
of stars.

\subsubsection{Molecular absorption}

The two detections of H$_2$ absorption in the MS with \emph{FUSE} 
toward Fairall\,9 and NGC\,3783 (Sembach et al.\,2001)
indicate that tidal gas streams around galaxies may 
typically host a widespread, cold gas phase that has 
a substantial absorption cross section. This scenario
is supported by the recent detection of H$_2$ absorption
in another, more distant circumgalactic tidal gas stream
beyond the Local Group (Crighton et al.\,2013).
These findings also remind us that 
the presence of H$_2$ in an intervening absorber with 
a complex velocity structure and a high neutral gas 
column density does not {\it necessarily} mean that one 
traces a gaseous disk of a galaxy. 

The relatively large 
absorption cross section of H$_2$ in neutral gas structures
{\it around} galaxies, as found in the Milky Way halo
(see Richter 2006 and references therein), can be explained
by the circumstance that the physical conditions in these
star-less gas clouds favour the formation of diffuse 
molecular structures even at relatively moderate gas 
densities. In particular, the relative low intensity 
of the local UV radiation field, the efficient process 
of H$_2$ line self-shielding, and the lack of the distructive 
influence of massive stars probably represent 
important aspects in this context.

H$_2$ observations in circumgalactic gas clouds not 
only provide important information on the physical and 
chemical conditions in these structures, they also are of
high relevance for our understanding of the physics
of molecular hydrogen in diffuse gas under conditions
that are very differnt from that in the local ISM
(see also Sembach et al.\,2001).
The transition from neutral to molecular gas, in particular,
represents one of the most important processes that govern the 
evolutionary state of galaxies at low and high redshift.
Detailed measurements of H$_2$ fractions and dust-depletion
patterns in (star-less) tidal streams at low redshift 
can help to constrain the critical formation rate of
molecular hydrogen in low-metallicity environments and
thus could be of great importance to better understand the
distribution and cross section of molecular gas in
and around high-redshift galaxies.

%

\section{Summary}

In this second paper of our ongoing series to study the Magellanic Stream
in absorption we have analyzed newly 
obtained optical and UV absorption-line data from \emph{HST}/COS and 
VLT/UVES together with archival \emph{FUSE} and H\,{\sc i} 21cm emission 
data from GASS and ATCA to study the chemical composition and the physical 
conditions in the Magellanic Stream in the direction of the quasar 
Fairall\,9. Our main results are the following:\\
\\
1. Metal absorption in the Magellanic Stream (MS) is detected 
in seven individual absorption components centered at 
$v_{\rm LSR}=+143,+163,+182,+194,+204$ and $+218$ km\,s$^{-1}$, 
indicating a complex internal velocity strucutre of the MS
in this direction. Detected ions, atoms and molecules in
the Stream include
C\,{\sc iv}, Si\,{\sc iv}, Si\,{\sc iii}, C\,{\sc ii}, 
C\,{\sc ii}$^{\star}$, Al\,{\sc ii}, Si\,{\sc ii}, 
S\,{\sc ii}, Ca\,{\sc ii}, Ti\,{\sc ii}, Fe\,{\sc ii}, 
Ti\,{\sc ii}, O\,{\sc i}, N\,{\sc i}, Na\,{\sc i}, and H$_2$.\\
\\
2. From the unsaturated S\,{\sc ii} absorption and a Cloudy
photoionization model we obtain an $\alpha$
abundance in the Stream of [S/H$]=-0.30\pm 0.04$ ($0.50$ solar),
which is substantially higher than that found in the Stream 
along the lines of sight toward NGC\,7469, RBS\,144, and NGC\,7714
(Fox et al.\,2010, 2013).
Unfortunately, the unresolved, saturated O\,{\sc i} $\lambda 1302.2$ 
line cannot be used to independently constrain the
$\alpha$ abundance in the MS toward Fairall\,9. Contrary to
sulfur, we measure a very low nitrogen abundance in the gas of 
[N/H$]=-1.15\pm 0.06$. The resulting [N/S] ratio is $-0.85$ dex,
which is very low compared to other environments with
similarly high $\alpha$ abundances.
The low [N/S] and high [S/H] ratios observed in the MS toward
Fairall\,9 suggest an abundance pattern that is dominated by
the $\alpha$ enrichment from massive stars and Type II supernovae,
while only very little primary nitrogen was deposited into
the gas when the Stream was separated from the 
Magellanic Clouds.\\
\\
3. The detection of very narrow Na\,{\sc i} and H$_2$ absorption 
(with $b\approx 2$ km\,s$^{-1}$) in the component at $v_{\rm LSR}=+188$ 
km\,s$^{-1}$ indicates the presence of a compact (pc-scale), cold gas stucure
in the MS along this sightline. From the analysis of the 
newly reduced archival \emph{FUSE} data of Fairall\,9  we measure a total 
molecular hydrogen column density of log $N$(H$_2)=17.93^{+0.19}_{-0.16}$, improving
previous results from Richter et al.\,(2001a). From the analysis 
of the H$_2$ rotational excitation we obtain a kinetic temperature
in the cold neutral gas phase of $T\approx 93$ K.
For the gas density we derive $n\approx 4-8$ cm$^{-3}$,
assuming that the H$_2$ gas is in a formation-dissociation
equilibrium. The resulting estimate for the thermal gas pressure
is $P/k\approx 750$ cm$^{-3}$\,K, thus in good agreement with values derived
from previous studies of the Stream. The detection of H$_2$ absorbing
structures in the MS, whose linear and angular sizes must be very
small ($\sim 1$ pc and $\sim 3$ arcseconds), indicates
that the neutral gas body of the Stream is highly clumped and
structured. We discuss the implications of physical and chemical 
inhomogeneities in circumgalactic gas structures on our understanding
of intervening QSO absorption-line systems. \\
\\
4. The relatively low column densities of Fe\,{\sc ii}, Ti\,{\sc ii},
Ni\,{\sc ii}, Al\,{\sc ii}, and Ca\,{\sc ii} indicate that the gas phase 
abundances of these elements are affected by dust depletion. We combine our
column-density measurements for these ions with a Cloudy 
photoionization model and derive dust
depletion values relative to sulfur of 
log $\delta$(Si$)= -0.27$,
log $\delta$(Fe$)= -0.56$,
log $\delta$(Ni$)\leq -0.51$,
log $\delta$(Ti$)= -1.32$, 
log $\delta$(Al$)= -0.62$, and
log $\delta$(Ca$)= 0$ to $-1.58$.
These depletion values are similar to those
found in warm, diffuse clouds in the lower
Milky Way halo (Savage \& Sembach 1996).\\
\\
5. Our study indicates that the enrichment history of the
Magellanic Stream as well as the physical conditions in the
Stream are more complex than previously known. 
The abundances and gas-to-dust ratios measured in the
Stream along the Fairall\,9 sightline are substantially
higher than what is found along other MS sightlines.
The high sulfur abundance in the gas possibly indicates a 
substantial $\alpha$ enrichment from massive stars 
in a region of enhanced star-formation $\sim 1-2$ Gyr ago
that then was stripped from the
Magellanic Clouds and incorporated into the MS 
before the gas could be enriched in nitrogen from
intermediate-mass stars.
Our findings are in line with the idea that the
metal-enriched filament in the Stream toward Fairall\,9 
originates in the LMC (Nidever et al.\,2008), but an
SMC origin is also possible (and slightly favoured by
the low nitrogen abundance in the gas).

\acknowledgments

We gratefully thank Gurtina Besla and Mary Putman for helpful 
discussions and Christian Br\"uns for providing the 
supporting ATCA 21cm data.
Support for this research was provided by NASA through 
grant HST-GO-12604 from the Space Telescope Science Institute, 
which is operated by the Association of Universities for Research 
in Astronomy, Incorporated, under NASA contract NAS5-26555.

%

\section*{REFERENCES}
\begin{footnotesize}

\noindent
Abgrall, H., \& Roueff, E. 1989, A\&A, 79, 313
\noindent
\\
Anders, E., \& Grevesse, N. 1989, Geochim. Cosmochim. Acta, 53, 197
\noindent
\\
Asplund, M., Grevesse, N., Jacques Sauval, A., \& Scott, P. 2009, ARA\&A, 47, 481
\noindent
\\
Ben Bekhti, N., Richter, P., Westmeier, T., \& Murphy, M.T. 2008,
A\&A, 487, 583
\noindent
\\
Ben Bekhti, N., Winkel, B., Richter, P., et al. 2011, A\&A, 542, 110
\noindent
\\
Besla, G., Kallivayalil, N., Hernquist, L., et al. 2007, ApJ, 668, 949
\noindent
\\
Besla, G., Kallivayalil, N., Hernquist, L., et al. 2010, ApJ, 721, L97
\noindent
\\
Besla, G., Kallivayalil, N., Hernquist, L., et al. 2012, MNRAS, 421, 2109
\noindent
\\
Bland-Hawthorn, J., \& Maloney, P. R. 1999, ApJ, 510, L33
\noindent
\\
Bland-Hawthorn, J., Sutherland, R., Agertz, O., \& Moore, B. 2007, ApJ, 670, L109
\noindent
\\
Bland-Hawthorn, J.,\& Maloney, P. R. 2002, in ASP Conf. Ser. 254, Extragalactic
Gas at Low Redshift, ed. J. S. Mulchaey \& J. T. Stocke (San Francisco: ASP),
267
\noindent
\\
Br\"uns, C., Kerp, J., Stavely-Smith, L., et al. 2005, A\&A, 432, 45
\noindent
\\
Crawford, I.A. 1992, MNRAS, 259, 47
\noindent
\\
Crighton, N. H. M., Bechtold, J., Carswell, R. F. et al. 2013, MNRAS, in press (astro-ph/1210.0905)
\noindent
\\
Connors, T. W., Kawata, D., Maddison, S. T., \& Gibson, B. K. 2004, PASA, 21, 222
\noindent
\\
Connors, T. W., Kawata, D., \& Gibson, B. K. 2006, MNRAS, 371, 108
\noindent
\\
Dalgarno, A., \& McCray, R. A. 1972, ARA\&A, 10, 375
\noindent
\\
de Boer, K. S., Richter, P., Bomans, D. J., Heithausen, A., \& Koornneef, J.
A\&A, 338, L5
\noindent
\\
Dekker, H., D'Odorico, S., Kaufer, A., Delabre, B., \& Kotzlowski,
H. 2000, SPIE, 4008, 534
\noindent
\\
Diaz, J. D., \& Bekki, K. 2011, ApJ, 413, 2015
\noindent
\\
Draine, B., \& Bertoldi, F. 1996, ApJ, 468, 269
\noindent
\\
Dufour, R.J. 1975, ApJ, 195, 315
\noindent
\\
Ferland, G. J., Korista, K. T., Verner, D. A. et al. 1998, PASP, 110, 761
\noindent
\\
Fontana, A., \& Ballester, P. 1995, ESO Messenger, 80, 37
\noindent
\\
Fox, A. J., Wakker, B. P., Savage, B. D., et al. 2005, ApJ, 630, 332
\noindent
\\
Fox, A.J., Wakker, B.P., Smoker, J.V., et al. 2010, ApJ, 718, 1046
\noindent
\\
Fox, A.J., Richter, P., Wakker, B.P. et al. 2013, ApJ, in press (astro-ph/1304.4240)
\noindent
\\
Gardiner, L.T. \& Noguchi, M. 1996, MNRAS, 278, 191
\noindent
\\
Gibson, B. K., Giroux, M. L., Penton, S. V., et al. 2000, AJ, 120, 1830
\noindent
\\
Green, J. C., Froning, C. S., Osterman, S., et al. 2012, ApJ, 744, 60
\noindent
\\
Harris, J., Zaritsky, D. 2004, AJ, 127, 1531
\noindent
\\
Harris, J., Zaritsky, D. 2009, AJ, 138, 1243
\noindent
\\
Heitsch, F. \& Putman, M. E. 2009, ApJ, 698, 1485
\noindent
\\
Henry, R. B. C., Edmunds, M. G, K\"oppen, J. 2000, ApJ, 541, 660
\noindent
\\
Henry, R. B. C. \& Prochaska, J. X. 2007, PASP, 119, 962
\noindent
\\
Hughes, J. P., Hayashi, I., \& Koyama, K. 1998, ApJ, 505, 732
\noindent
\\
Irwin, M. J., Kunkel, W. E., Demers, S. 1985, Nature, 318, 160
\noindent
\\
Jenkins, E. B., Bowen, D. V., Tripp, T. M., \& Sembach, K. R. 2005,
ApJ, 623, 767
\noindent
\\
Kalberla, P. M. W., Burton, W. B., Hartmann, D., et al. 2005, 
A\&A, 440, 775
\noindent
\\
Kalberla, P. M. W., McClure-Griffiths, N. M., Pisano, D. J., et al.
2010, A\&A, 521, 17
\noindent
\\
Kallivayalil, N., van der Marel, R. P., Alcock, C., et al. 2006a,
ApJ, 638, 772
\noindent
\\
Kallivayalil, N., van der Marel, R. P., \& Alcock, C. 2006b, ApJ,
652, 1213
\noindent
\\
Kallivayalil, N., van der Marel, R. P., Besla, G., Anderson, J., \&  Alcock, C. 2013,
ApJ, 764, 161
\noindent
\\
Keenan, F. P., Lennon, D. J., Johnson, C. T., \& Kingston, A. E. 1986,
MNRAS, 220, 571
\noindent
\\
Kobayashi, C., Umeda, H., Nomoto, K., Tominaga, N., \& Ohkubo, T. 2006,
ApJ, 653, 1145
\noindent
\\
Koerwer, J.F. 2009, AJ, 138, 1
\noindent
\\
Kriss, G. A. 2011, \emph{COS} Instrument Science Report, 1
\noindent
\\
Lehner, N. 2002, ApJ, 578, 126
\noindent
\\
Lehner, N., Wakker, B.P., \& Savage, B.D. 2004, ApJ, 615, 767
\noindent
\\
Lehner, N., Howk, J.C., Keenan, F.P., \& Smoker, J.V. 2008, ApJ 678, 219
\noindent
\\
Lu, L., Savage, B. D., \& Sembach, K. R. 1994, ApJ, 437, L119
\noindent
\\
Mastropietro, C., Moore, B.,Mayer, L.,Wadsley, J., \& Stadel, J. 2005, MNRAS,
363, 509
\noindent
\\
Mathewson, D. S., Ford, V. L., Schwarz, M. P., \& Murray, J. D. 1979, IAUS, 84, 547
\noindent
\\
Melnick, J. 1985, A\&A, 153, 235
\noindent
\\
Morton, D. C., \& Dinerstein, H. L. 1976, ApJ, 204, 1
\noindent
\\
Morton, D. C. 2003, ApJS, 149, 205
\noindent
\\
McClure-Griffiths, N. M., Pisano, D. J., Calabretta, M. R., et al. 2009, 
ApJS, 181, 398
\noindent
\\
Nidever, D. L., Majewski, S. R., \& Burton, W. B. 2008, ApJ, 679, 432
\noindent
\\
Nidever, D. L., Majewski, S. R., Burton, W. B., \& Nigra, L. 2010,
ApJ, 723, 1618
\noindent
\\
Pagel, B. E. J., \& Tautvaisiene, G. 1998, MNRAS, 299, 535
\noindent
\\
Pettini, M., Ellison, S. L., Bergeron, J., \& Petitjean, P. 2002,
A\&A, 391, 21
\noindent
\\
Pettini, M., Zych, B. J., Steidel, C. C., \& Chaffee, F. H. 2008,
MNRAS, 385, 2011
\noindent
\\
Putman, M. E., Staveley-Smith, L., Freeman, K. C., Gibson, B. K., \& Barnes, D. G. 2003,
ApJ, 586, 170
\noindent
\\
Ribaudo, J., Lehner, N., Howk, J. C., et al. 2011,
ApJ, 743, 207
\noindent
\\
Richter, P., Widmann, H., de Boer, K. S. et al. 1998, A\&A, 338, L9
\noindent
\\
Richter, P. 2000, A\&A, 359, 1111
\noindent
\\
Richter, P., Sembach, K. R., Wakker, B. P., \& Savage, B. D., 2001a,
ApJ, 562, L181
\noindent
\\
Richter, P., Savage, B. D., Wakker, B. P., Sembach, K. R., \& Kalberla, P. M. W.
2001b, ApJ, 549, 281
\noindent
\\
Richter P., Wakker B. P., Savage B. D., \& Sembach K. R. 2003, ApJ, 586, 230
\noindent
\\
Richter, P., Westmeier, T., \& Br\"uns 2005,  A\&A, 442, L49
\noindent
\\
Richter, P. 2006, Rev.\,Mod.\,Astron., 19, 31
\noindent
\\
Richter, P., Charlton, J. C., Fangano, A. P. M., Ben Bekhti, N.,
\& Masiero, J. R. 2009, ApJ, 695, 1631
\noindent
\\
Richter, P., Krause, F., Fechner, C., Charlton, J. C.,
\& Murphy, M. T. 2011, A\&A, 528, A12
\noindent
\\
Russell, S. C., \& Dopita, M. A. 1990, ApJ, 74, 93
\noindent
\\
Russell, S. C., \& Dopita, M. A. 1992, ApJ, 384, 508
\noindent
\\
Savage, B. D., Drake, J. F., Budich, W., \& Bohlin, R. C. 1977, ApJ,
216, 291
\noindent
\\
Savage, B. D., \& Sembach, K. R. 1991, ApJ, 379, 245
\noindent
\\
Savage, B. D. \& Sembach, K. R. 1996, ARA\&A, 34, 279
\noindent
\\
Sembach, K. R., Howk, J. C., Savage, B. D., \& Shull, J. M. 2001,
AJ, 121, 992
\noindent
\\
Songaila, A. 1981, ApJ, 243, L19
\noindent
\\
Spitzer, L. 1978, {\it Physical processes in the interstellar medium}, 
(New York Wiley-Interscience)
\noindent
\\
Thom, C., Werk, J. K., Tumlinson, J., et al. 2011, ApJ, 736, 1
\noindent
\\
Tumlinson, J., Shull, J. M., Rachford, B. L., et al. 2002, ApJ, 566, 857
\noindent
\\
Tumlinson, J., Werk, J. K., Thom, C., et al. 2011, ApJ, 733, 111
\noindent
\\
Wakker, B. P., Howk, J. C., Savage, B. D., et al. 1999, Nature, 402, 388
\noindent
\\
Wakker, B. P., Oosterloo, T. A., \&  Putman, M. E. 2002, AJ, 123, 1953
\noindent
\\
Wakker, B.P. 2006, ApJS, 163, 282
\noindent
\\
Wakker, B. P., York, D. G., Howk, J. C., et al.\,2007, ApJ, 670, L113
\noindent
\\
Wakker, B. P., York, D. G., Wilhelm, R., et al. 2008, ApJ, 672, 298
\noindent
\\
Wakker, B. P., Lockman, F. J., \& Brown, J. M. 2011, ApJ, 728, 159
\noindent
\\
Wannier, P., \& Wrixon, G. T. 1972, ApJ, 173, L119
\noindent
\\
Weiner, B. J., \& Williams, T. B. 1996, AJ, 111, 1156
\noindent
\\
Weisz, D. R., Doplhin, A. E., Skillmann, E. D., et al.\,2013, MNRAS, 431, 364
\noindent
\\
Welty, D. E., Morton, D. C., \& Hobbs, L.M. 1996, ApJS, 106, 533
\noindent
\\
Welty, D. E., Lauroesch, J. T., Blades, J. C., Hobbs, L.M., \& York, D. G. 1997,
ApJ, 489, 672
\noindent
\\
Welty, D. E., Federman, S. R., Gredel, R., Thorburn, J. A., \& Lambert, D. L. 2006,
ApJS, 165, 138
\noindent
\\
Wolfire, M. G., McKee, C. F., Hollenbach, D., \& Tielens, A. G. G. M.
1995, ApJ, 453, 673
\noindent
\\
York, D. G., Songaila, A., Blades, J. C., et al. 1982, ApJ, 255, 467
\noindent
\\
Zech, W. F., Lehner, N., Howk, J. C., Dixon, W. V. D., \& Brown, T. M. 2008, ApJ,
679, 460

\end{footnotesize}

%

\newpage

\appendix

\section{H\,{\sc i} Ly\,$\alpha$ absorption}

%

\begin{figure}[t!]
\epsscale{1.0}
\plotone{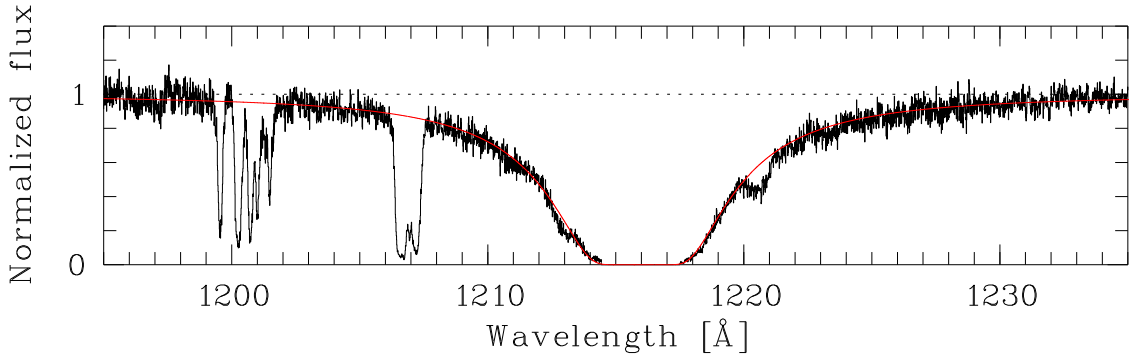}
\caption{H\,{\sc i} Ly\,$\alpha$ absorption in the COS spectrum
of Fairall\,9 in the wavelength range between $1195$ and $1235$ \AA.
The red solid line indicates a two-component absorption model
with a Milky Way H\,{\sc i} column density of
log $N$(H\,{\sc i}$)=20.25$ and a MS H\,{\sc i} column density of
log $N$(H\,{\sc i}$)=19.95$.
}
\end{figure}

%

In Fig.\,10 we show the normalized COS spectrum in the wavelength range 
between $1195$ and $1235$ \AA. Strong H\,{\sc i} Ly\,$\alpha$ 
absorption centered at $1216$ \AA\, from neutral gas in the 
Milky Way (MW) and in the Magellanic Stream with well-defined
damping wings is visible. The 21cm data from various 
observations and instruments (Sect.\,3.3; Gibson et al.\,2000)
indicate that the Milky Way disk gas dominates the neutral-gas
column density along this sightline.

The strength and shape of the Ly\,$\alpha$ absorption toward
Fairall\,9 supports this conclusion, but the relative
contribution of MW and MS cannot be tightly constrained from a 
Voigt-profile fit of the Ly\,$\alpha$ absorption. The red solid
line in Fig.\,10 shows a two-component absorption model
with a MW H\,{\sc i} column density of
log $N$(H\,{\sc i}$)=20.25$ and a MS H\,{\sc i} column density of
log $N$(H\,{\sc i}$)=19.95$. The Ly\,$\alpha$ absorption
profile is fully consistent with the value of log $N$(H\,{\sc i}$)=19.95$
in MS, as indicated by the GASS 21cm data. However, the Ly\,$\alpha$
profile is also consistent with MS column densities in the range
log $N$(H\,{\sc i}$)=19.75-20.15$, if the MW contribution is
modified accordingly. In conclusion, the Ly\,$\alpha$ absorption
does not provide constraining limits on the H\,{\sc i} column
density in the Magellanic Stream toward Fairall\,9 that could
help to minimize the systematic errors for the determination
of log $N$(H\,{\sc i}), as discussed in Sect.\,3.3.

\section{Absorption modeling of UV lines}

In Sect.\,2.5 we have outlined the method that we have used to
model the UV lines of the low ions in the COS spectrum of Fairall\,9
to infer column densities of these species. We here present 
additional information on the modeling procedure to provide
an insight into the allowed parameter range for $(N_i,b_i)$ and
$N_{\rm tot}$ for each ion. 

In Fig.\,11 we show velocity plots of the low ions from 
synthetic spectra that have been generated by us based on
different input models, in which we have systematically varied 
the parameters $N_i$ and $b_i$. The resulting model profiles 
are compared to the best-fitting model presented in Fig.\,3 
and Table\,1.

In the left panel of Fig.\,11 (a), the best-fitting model is indicated
with the black solid line, while the lines in red, green, and blue 
show models, in which all column densities $N_i$ in the individual
subcomponents have been increased by $0.3$, $0.5$, and $1.0$ dex,
respectively. The velocity structure and the $b$ values remain
unchanged. In the middle panel of Fig.\,11 (b), the column densities 
$N_i$ in the individual subcomponents have been decreased by $0.3$, 
$0.5$, and $1.0$ dex (red, green, blue), again without changes in 
$v_i$ and $b_i$. These two figures indicate that the shapes of the 
strong and saturated lines from C\,{\sc ii}, O\,{\sc i} and Si\,{\sc ii}
are basically not affected by the change in column density. In contrast, the
absorption depths of the weaker lines of N\,{\sc i}, Al\,{\sc ii}, Si\,{\sc ii}, 
S\,{\sc ii}, and Fe\,{\sc ii} are {\it strongly} affected by the increase or
decrease of the column densities with respect to the reference model
(black), demonstrating that these lines represent sensitive diagnostics to 
constrain $N_i$  and $N_{\rm tot}$ for these ions in the Stream.

In the right panel of Fig.\,11 (c) we show a model in which we have
increased the Doppler parameters $b_i$ for the individual subcomponents
by factors of $1.2$ (red), $1.5$ (green), and $2.0$ (blue)
with respect to the reference model (black), while the column densities and 
the overall velocity structure remain unchanged. Because of the
relatively low spectral resolution of COS compared to the
absolute values of $b_i$, the changes in $b_i$
are irrelevant for the shape of the weak, unsaturated lines,
while the strong, saturated lines mildly grow in strength in
their absorption wings.
Therefore, even if we would underestimate the values of $b_i$ 
for the singly-ionized species Al\,{\sc ii}, Si\,{\sc ii},
S\,{\sc ii}, and Fe\,{\sc ii} when adopting the values 
from Ca\,{\sc ii}, as one could argue because these ions
may trace a slightly different (more extended) gas phase,
our column-density estimate would not be affected at all 
by this systematic error.

These figures, together with Fig.\,2, underline that the total gas 
column densities of N\,{\sc i}, Al\,{\sc ii}, Si\,{\sc ii},
S\,{\sc ii}, and Fe\,{\sc ii} in the MS, as listed in Table\,1, are 
well constrained by the COS data and our absorption model.

%

\begin{figure}[t!]
\epsscale{0.8}
\plotone{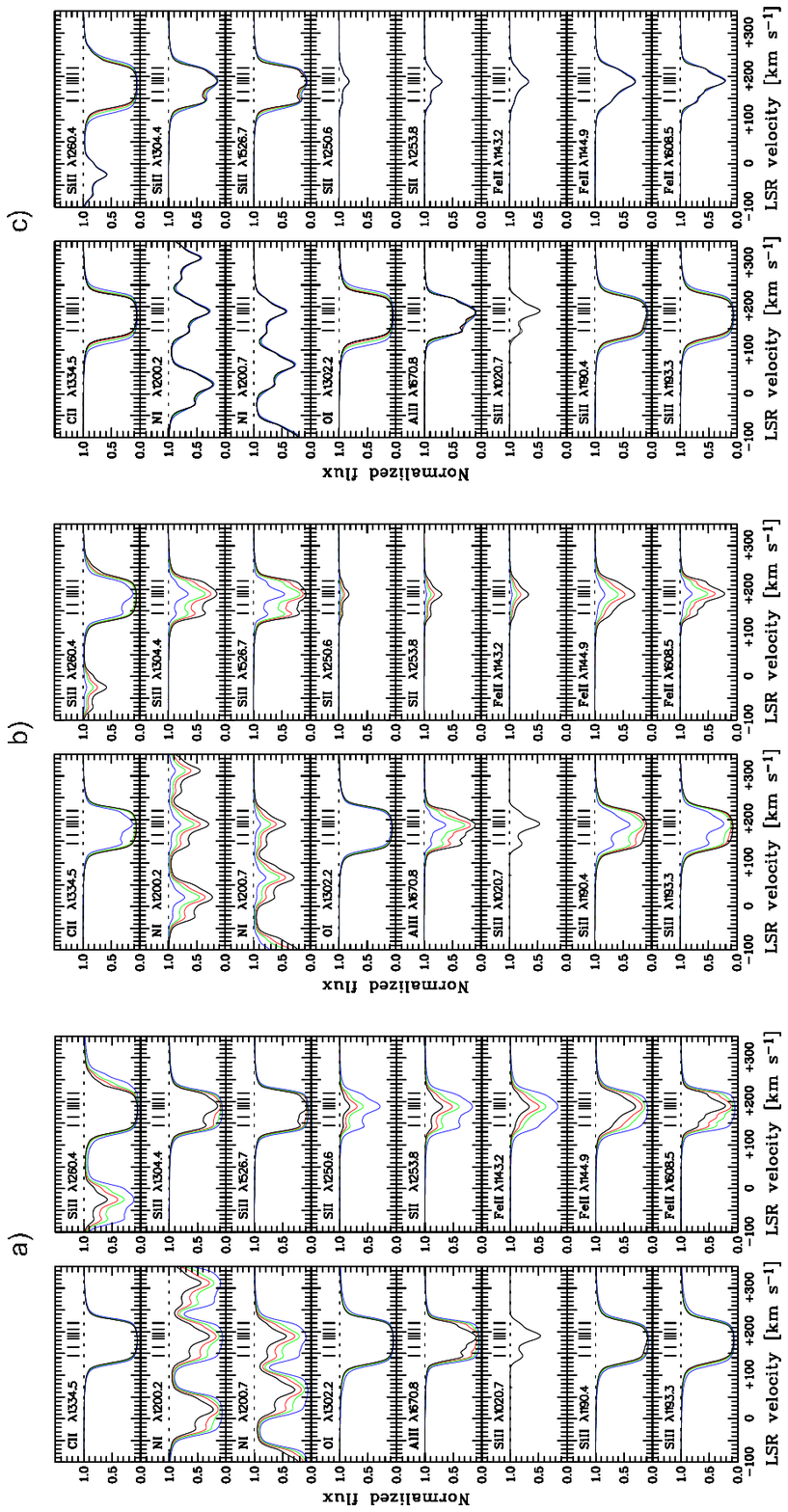}
\caption{Velocity profiles of low ions, as derived from synthetic spectra
from different input models for the gas absorption in the Magellanic Stream
toward Fairall\,9. The best-fitting model is indicated with the black solid 
line. The lines in red, green, and blue show test models, in which the
input parameters $N_i$,  $b_i$ have been varied systematically to
study the resulting absorption profiles. Details on these test
models are given in Appendix B.
}
\end{figure}

%

\end{document}